\documentclass[letter]{aa}
\usepackage{amsmath}
\usepackage{txfonts}
\usepackage{graphicx}
\usepackage{natbib}
\bibpunct{(}{)}{;}{a}{}{,}
\usepackage{dcolumn}

\begin{document}

\title{The IDV source J\,1128+5925, a new candidate for annual modulation?}
\author{K.\,\'E. Gab\'anyi\inst{1,2} 
\and  N. Marchili\inst{1} 
\and T.\,P. Krichbaum\inst{1} 
\and S. Britzen\inst{1} 
\and L. Fuhrmann\inst{1} 
\and A. Witzel\inst{1} 
\and J.\,A. Zensus\inst{1} 
\and P.\,M\"uller\inst{1} 
\and X. Liu\inst{3} 
\and H.~G. Song\inst{3} 
\and J.~L. Han\inst{4} 
\and X.~H. Sun\inst{4}}

\institute{Max-Planck-Institut f\"ur Radioastronomie (MPIfR), Auf dem H\"ugel 69, 53121,
Bonn, Germany 
\and Hungarian Academy of Sciences (HAS) Research Group for Physical Geodesy and Geodynamics, Budapest, Hungary 
\and Urumqi Observatory,the National Astronomical Observatories, the Chinese Academy of Sciences, Urumqi 830011, PR China 
\and National Astronomical Observatories, Chinese Academy of Sciences, Beijing 100012, PR China}
\offprints{K.\,\'E. Gab\'anyi, \email{gabanyik@sgo.fomi.hu}}
\date{Received / Accepted }

\abstract
{Short time-scale radio variations of compact extragalactic radio sources, 
known as IntraDay Variability (IDV), can be explained in at least some sources 
by a source-extrinsic effect, 
in which the variations are interpreted as scintillation of radio waves caused by 
the turbulent interstellar medium of the Milky Way. One of the most convincing 
observational arguments in favour of propagation-induced variability is the 
so called ``annual modulation'' of the characteristic variability time-scale,
which is due to the orbital motion of the Earth. So far there are only two 
sources known, which show such a well-defined seasonal cycle, 
a few more sources with fewer data can be regarded as possible candidates for
this effect.
However, source-intrinsic effects, such as structural variations, can also 
cause the observed changes of the variability time-scale. Data for the new, 
recently discovered and highly variable IDV source J\,1128+5925 are presented.}
{We study the frequency and time dependence of the IDV in this 
compact quasar. We measure the characteristic variability time-scale of 
the IDV throughout the year, and analyze whether the observed changes 
in the variability time-scale are consistent with annual modulation.
Assuming a radio wave propagation effect as origin, we are able to constrain
some physical properties (such as distance, scattering-strength and possible
anisotropy) of the ``plasma'' screen, which may cause the 
scintillation.}
{We monitored the flux density variability of J\,1128+5925 with dense time 
sampling between 2.7 and 10.45\,GHz. We observed with
the 100\,meter Effelsberg radio telescope of the Max-Planck-Institut f\"ur Radioastronomie
(MPIfR) at 2.70\,GHz, 4.85\,GHz, and 10.45\,GHz, and with the 25\,meter Urumqi 
radio telescope (China) at 4.85\,GHz. From ten observing sessions, each of several
days in duration and performed between 2004 -- 2006, we 
determine the variability characteristics and time-scales
and investigate these in view of possible scintillation and annual modulation.}
{The observed pronounced changes of the variability time-scale 
of J\,1128+5925 are modelled with an anisotropic annual modulation model. 
The observed frequency dependence of the variation is in good agreement 
with the prediction from interstellar scintillation. Adopting a simple model
for the annual modulation model and using also the 
frequency dependence of the IDV, we derive a lower limit to the distance of the
scattering screen and an upper limit to the scintillating source size. The 
latter is found to be consistent with the measured core size from Very 
Long Baseline Interferometry (VLBI).}
{}


\authorrunning{K.\,\'E. Gab\'anyi et al.}
\titlerunning{J\,1128+5925: Annual Modulation in a new IDV source}

\maketitle

\section{Introduction} \label{intro}

The so-called IntraDay Variability (IDV) of compact flat-spectrum radio 
sources (quasars and blazars) was discovered in the mid-eighties by  
\cite{idv_discovery} and \cite{idv_discovery2}. If interpreted as being 
source intrinsic, the short variability time-scale (of 1--2 days) 
would imply - through causality and light-travel time arguments - extremely small 
sizes of the emitting regions ($\mu$arcsecond-scale), leading to
very large apparent brightness temperatures 
(in the range of  $10^{16} \text{\,K to } 10^{21}$\,K).
This is far in excess of the inverse-Compton limit 
of $\sim 10^{12}$\,K \citep{compton} and would require with the assumption
of a spherical geometry excessively large Doppler boosting factors 
($\delta$\,$>>$\,$50$). In somewhat special, non-spherical source geometries 
\citep[such as cylindrical shock-in-jet models 
e.g.][]{qian_0716, qian_injection, qian_0917, spada}, the requirements for
the Doppler factor would be less extreme. Alternatively,
coherent processes and collective plasma emission 
\citep{coherent_benford, coherent_lesch}, or a short time intrinsic violation
of the inverse Compton limit \citep{slysh} are processes used to explain IDV.

While so far no direct other observational evidence exists for the assumption
of such extremely large Doppler factors, the interpretation of IDV via
source extrinsic effects appears to be very successful, and in particular
explains the very fast (hourly) variations seen in some
extremely rapid IDV sources \citep[e.g.:][]{dt}. 
In this source-extrinsic interpretation \citep[e.g.:][]{rick_uj}, the observed IDV is completely
caused by interstellar scintillation (ISS) of radio waves passing through the 
turbulent plasma of the Milky Way (propagation effect). However, the 
emission and source structure of compact extragalactic radio sources 
(quasars, blazars) is known to vary 
also intrinsically on almost every time scale, from minutes to years. Thus it
is very difficult to discriminate between a source intrinsic and a source
extrinsic origin of IDV at radio-bands.

One of the most convincing arguments in favour of a source extrinsic 
explanation of IDV is the so called annual 
modulation of the variability time-scale. An annual cycle in the 
observed variability time-scale is caused by the variations 
of the relative velocity vector between the scattering 
screen and  the velocity of the Earth as the Earth orbits 
around the Sun \citep{annual1819}.
Such seasonal cycles are seen in two IDV sources: 
J\,1819+3845 \citep{annual1819} and PKS\,1257-326
\citep{time_delay3}. In a few of other IDV sources, 
such as 
PKS\,1519-273 \citep{annual1519_1}, 
B\,0917+624 \citep{0917annual1, 0917annual2}, 
PKS\,0405-385 \citep{0405annual} and 
S5\,0954+658 (Fuhrmann et al. in prep., 2007),
the observed variability timescales do not show such a clear effect,
and a possible seasonal pattern is either not present or is smeared out.
For two sources, B\,0917+624 \citep{cease_0917} and PKS\,0405-385, so called
episodic IDV \citep[e.g.][]{0405annual} is observed, where previously observed
pronounced IDV, either temporarily disappears (PKS\,0405-385), or even ceases
(B\,0917+624). This makes it difficult to prove beyond  doubt the existence 
of any annual modulation pattern. We note that such episodic IDV can also be due
to changes of the source structure (i.e. expansion of previously scintillating
structure components) or due to changes in the properties of the scattering
plasma.

The other observational argument for an extrinsic explanation of IDV is 
the ``time-delay measurements'', where
a time delay is measured between the IDV pattern arrival times at two telescopes.
Such time delay of the variability pattern was observed in PKS\,0405-385 
between the Australia Telescope  Compact Array (ATCA) and the Very Long Array (VLA) \cite{timedelay_0405}, in J\,1819+3845 between 
the Westerbork Synthesis Radio Telescope (WSRT, Netherlands) and the VLA \citep{timedelay_1819}, and in PKS\,1257-326 
between the ATCA and the VLA \citep{bignall_newest}. These 
delay measurements, when combined with the observations of the annual modulation
cycle yield constraints to the velocity, geometry and the distance of the plasma
causing the scintillation.  
The time delay measurements are feasible only for rapidly varying IDV sources,
where the variability time scale is significantly shorter than the duration
of the observing interval. These so called fast scintillators
have characteristic variability time-scales of hours or less than an hour. 
In contrast, the variability time-scales of classical type II IDV sources 
\citep{idv_discovery2} are much longer, typically 
two days, and so time-delay measurements cannot be conducted for these sources.

To establish convincingly an annual modulation pattern, the source has to be 
observed regularly during the course of the year and 
for several years. To facilitate the measurement of a characteristic 
variability time-scale, the duration of an individual IDV session must be 
long enough, so that several  ``scintillation'' events (``scintles'') can be observed.
Within the scheduling constraints of large observatories, the regular observations
of fast scintillators, which vary on time-scales of a few hours or less,
is easier to perform than for the slower, classical type II IDV sources. For
these sources throughout the year regularly performed IDV observations (each lasting at least
3-4 days) cannot be performed with an oversubscribed telescopes,
like the MPIfR 100\,meter Effelsberg radio telescope. The search for annual modulation
in classical type II sources has to invoke other radio telescopes, which can
provide more observing time.

In this paper, we present ten epochs of IDV observations for a recently
discovered and rapidly variable radio source
GB6\,J\,1128+5925 (henceforth referred to as J\,1128+5925): six epochs 
measured with the MPIfR 100\,meter Effelsberg radio telescope at wavelengths
2.8\,cm, 6\,cm and 11\,cm, and six epochs measured with the 25\,meter radio telescope 
in Urumqi (China) at 6\,cm wavelength. These observations were aimed to
study the IDV characteristics and to measure the IDV time-scale throughout the
year.

J\,1128+5925 is a radio quasar at a redshift $z=1.795$
\citep{spectrum}, the 2000.0 epoch equatorial coordinates of J\,1128+5925 are
$\alpha_\text{2000.0}=11^\text{h}28^\text{m}13.3406^\text{s}$ and $\delta_\text{2000.0}=+59^\circ25'14.798''$ \citep{position}.  

\begin{figure}
 \begin{center}
  \includegraphics[width=\columnwidth]{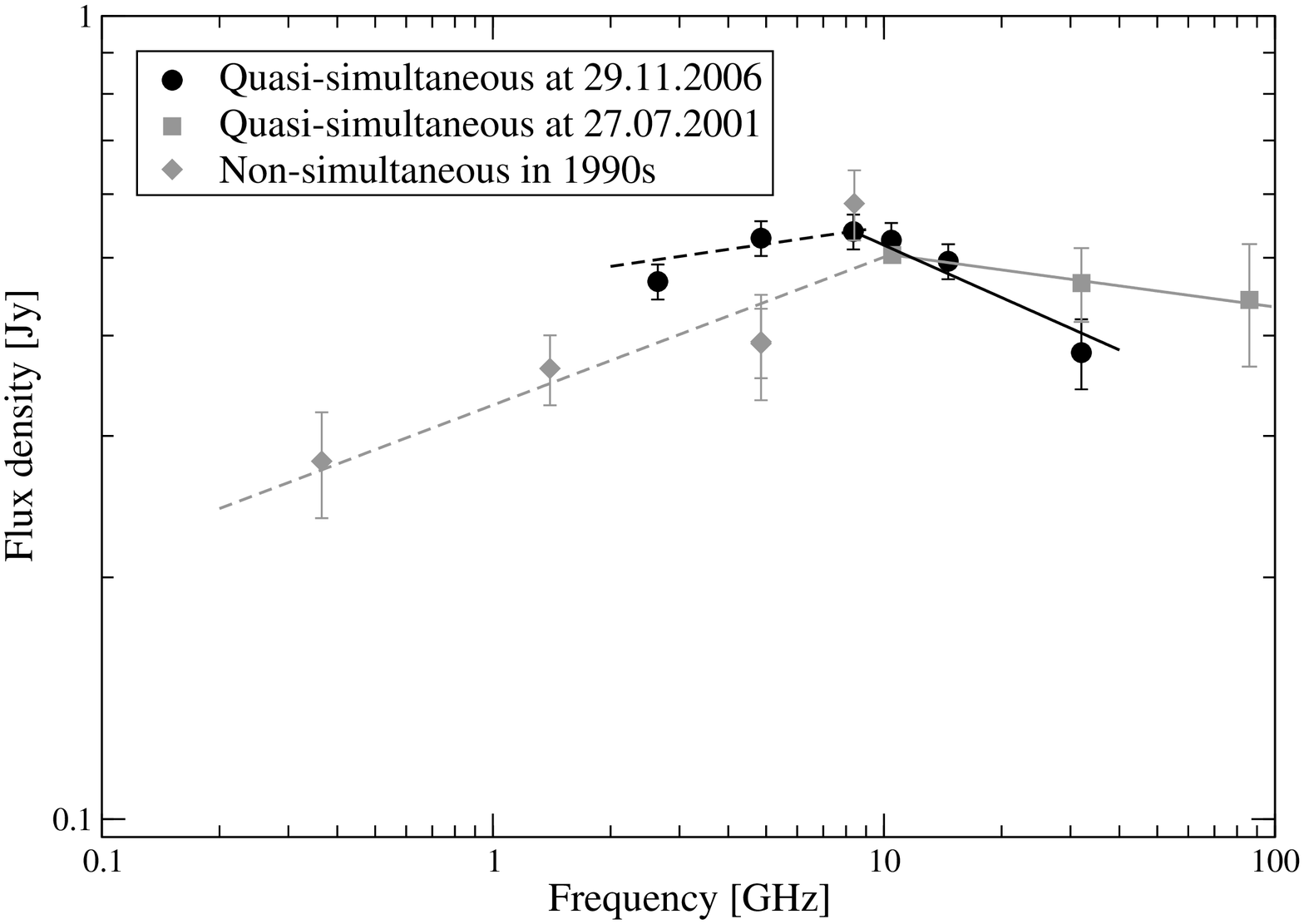}
  \caption{The radio spectrum of J\,1128+5925. Grey diamonds summarize flux density measurements from \cite{flux8,GB6,flux5b,flux1.4} and \cite{flux_texas}. The grey dashed line represents the fit to these data points, the corresponding spectral index $\alpha_{0.365\text{\,GHz}}^{8.4\text{\,GHz}}$ is $0.18 \pm 0.07$. Filled grey squares denote for quasi-simultaneous flux-density measurements with the MPIfR 100\,meter Effelsberg radio telescope and the IRAM Plateau de Bure interferometer, observed on 22th July 2001. The solid, grey line represents the fit to these data points, the corresponding spectral index $\alpha_{10.5\text{\,GHz}}^{86\text{\,GHz}}$ is $-0.07 \pm 0.015$. The filled black-circles show quasi-simultaneous flux-density measurements obtained with the Effelsberg telescope on 29th November 2006. The dashed line is a fit to the data up to 8.35\,GHz; the corresponding spectral index $\alpha_{2.64\text{\,GHz}}^{8.35\text{\,GHz}}$ is $0.07\pm0.04 $. The solid black line is a fit to the data between 8.35\,GHz and 32\,GHz; the corresponding spectral index $\alpha_{8.35\text{\,GHz}}^{32\text{\,GHz}}$ is $-0.22\pm0.04$. }
\label{fig:big_spektrum}
 \end{center} 
\end{figure}

The radio spectrum of J\,1128+5925 was observed at several dates and is shown
in Fig. \ref{fig:big_spektrum}. The spectrum peaks
between 8\,GHz and 10\,GHz. Between 0.365\,GHz and 8.4\,GHz, a spectral index
based on non-simultaneous data from the literature
(\cite{flux8,GB6,flux5b,flux1.4} and \cite{flux_texas}) of
$\alpha_{0.365\text{\,GHz}}^{8.4\text{\,GHz}}=0.18 \pm 0.07$
\footnote{The spectral index is defined as: $S\sim \nu^{+\alpha}$} is measured.
Quasi-simultaneous observations performed with the Effelsberg radio telescope
and the Plateau de Bure interferometer of the Insititut de Radio Astronomie
Millim\'etrique (IRAM) in 2001 yield a 
flat high frequency spectrum, with
spectral index $\alpha_{10.5\text{\,GHz}}^{86\text{\,GHz}}=-0.07 \pm 0.01$.
Quasi-simultaneous measurements performed in 2006 show some 
spectral variability when compared to earlier observations, now
yielding $\alpha_{2.64\text{\,GHz}}^{8.35\text{\,GHz}}=0.07 \pm 0.04$
below the turnover frequency, and
$\alpha_{8.35\text{\,GHz}}^{32.0\text{\,GHz}}=-0.22 \pm 0.04$ above this turnover.
The spectral shape and variability is consistent with an inhomogeneous synchrotron
self-absorbed compact radio source, which undergoes some long-term variability.

In the following we describe the organization of the paper. In Sect. \ref{obs}, we 
summarize the observational details, in Sect. \ref{analy} we describe the 
basic methods used in the variability
analysis of the data, in Sect. \ref{res} we present the results and discuss them in
the framework of the annual modulation scenario. Sect. \ref{concl} contains
a summary and conclusion.

The following cosmological parameters were used throughout this paper:  
$H_{0}=71\,\text{km\,s}^{-1}\text{Mpc}^{-1}\text{, } \Omega_{\text{matter}}=0.27 \text{ and } \Omega_{\text{vac}}=0.73$.

\section{Observations and data reduction} \label{obs}

The observations were performed with the MPIfR 100\,meter Effelsberg radio telescope at
centre frequencies 2.70\,GHz (11\,cm), 4.85\,GHz (6\,cm) and 10.45\,GHz (2.8\,cm), and with 
the Urumqi radio telescope at 4.85\,GHz. The 6\,cm receiver, a new backend and a new 
driving program for the Urumqi telescope was provided by the MPIfR.
A detailed technical description of the 6\,cm receiver system and the radio 
telescope are given in e.g. \cite{urumqi_tel}.

J\,1128+5925 has been observed at 10 different epochs, which are inhomogenously distributed
over a $1.5$ year interval between December 2004 and July 2006. In two epochs (December 2005 and 
April 2006), the Urumqi and Effelsberg telescopes observed quasi-simultaneously, to allow
for checks of the consistency of the flux-density variability seen with the two telescopes. 
The details of the observations are summarized in Table \ref{tab:obs1128}.

\begin{table}
\caption{\label{tab:obs1128}Summary of IDV observations of J\,1128+5925. The
  table lists the observing dates (Col. 1), the observing radio telescopes
  (Col. 2, 'E' for Effelsberg, 'U' for Urumqi), the observing frequencies (Col. 3), the total observing time (Col. 4), and the 
mean time sampling for the flux measurements of J\,1128+5925 (Col. 5). Letters
shown in the last column (Col. 5) refer to the corresponding variability curve, 
shown in Fig. \ref{fig:lcs}.}
\centering
\begin{tabular}{ccccccc}
\hline 
\hline
Epoch & R.T. & \multicolumn{1}{c}{$\nu$} & \multicolumn{1}{c}{Duration} & \multicolumn{1}{c}{Sampling} & \\
 & & \multicolumn{1}{c}{[GHz]} & \multicolumn{1}{c}{[h]} & \multicolumn{1}{c}{points/hr} \\
\hline
25-31.12.2004 & E & 4.85 & 125 & 0.2 & (a) \\
13-16.05.2005 & E & 4.85 & 67 & 0.8 & (b) \\
13-16.05.2005 & E & 10.45 & 35 & 0.8 & \\
14-17.08.2005 & U & 4.85 & 67 & 0.4 & (c) \\
16-19.09.2005 & E & 4.85 & 64 & 1.4 & (d) \\
27-31.12.2005 & U & 4.85 & 87 & 0.5 & (e) \\
29-30.12.2005 & E & 2.70 & 29 & 1.0 & \\
29-30.12.2005 & E & 4.85 & 30 & 2.0 & (e) \\
29-30.12.2005 & E & 10.45 & 30 & 1.4 & \\
10-12.02.2006 & E & 2.70 & 43 & 1.1 & \\
10-12.02.2006 & E & 4.85 & 43 & 2.0 & (f) \\
15-18.03.2006 & U & 4.85 & 71 & 0.6 & (g) \\
28.04-02.05.2006 & E & 2.70 & 86 & 1.7 & \\
28.04-02.05.2006 & E & 4.85 & 86 & 1.4 & (h) \\
28.04-02.05.2006 & U & 4.85 & 94 & 1.1 & (h) \\
10-13.06.2006 & U & 4.85 & 77 & 0.9 & (i) \\
14-18.07.2006 & U & 4.85 & 95 & 0.8 & (j) \\
\hline
\end{tabular}
\end{table}

At both telescopes,
the flux density measurements were performed using the cross-scan technique, 
in which the telescope was repeatedly moved over the source position 
in azimuth (Az) and in elevation (El) direction. Four to eight of these slews (sub-scans)
in each driving direction formed a scan, which, because of the point-like
brightness distribution of the source, is of Gaussian shape.
After subtraction of a linear baseline, which removes residual
system temperature drifts during a sub-scan, a
Gaussian curve was fitted to each of the resulting slices across the source.
The peak amplitude of the Gaussian yields a measure of the antenna temperature 
and of the source strength. The on-source integration time from the 4--8
subscans was found to be adequate to obtain flux-density measurements
with a signal-to-noise ratio of SNR $> 10^4$ at all observing frequencies, 
which provides sufficient measurement accuracy.

The Effelsberg receivers at 6\,cm and 2.8\,cm are  dual-horn beam-switch systems,
which allow to observe simultaneously the source and nearby blank sky 
(meaning, without the source). The immediate subtraction of the two signals
removes the atmospheric fluctuations from the source signal.
For a detailed description see e.g. \cite{idv_discovery2} and \cite{reduc_idv}.

In contrast to Effelsberg, the 6\,cm at Urumqi measured with a single horn only, 
therefore the fluctuating weather effects could not be subtracted. This reduces
the measurement accuracy when the weather is not optimal, and in the presence
of rain or rapidly moving clouds led to data, which could not be used for the
further data analysis (data removed by manual data editing).

In the next step, the amplitudes of the Gaussian fits to the individual
sub-scans were independently averaged in each driving direction (Az, El),
and small corrections for the residual telescope pointing errors 
were applied. In the last step the so corrected amplitudes from
both driving directions were averaged together, yielding a single 
antenna temperature measurement for each source and scan.

The next step of the data reduction is to correct for the systematic 
elevation dependent gain variations. Elevation dependent effects mainly originate 
from the gravitational deformation of the dish, which
causes variations of the telescope gain with changing elevation. A polynomial correction
was determined using the combined gain-elevation curves obtained from the 
non-variable calibrator sources.

After this, the time-dependent gain variations must be corrected.
These time-dependent effects are mainly caused by the changing weather 
conditions during the observations, by long-term gain fluctuations of the
receivers and from thermal gradients in the telescope. 
These effects can be corrected using a gain-time transfer function derived
from measurements of several secondary calibrator sources
(e.g. B\,0836+710 and B\,0951+699), which are observed
with the same duty cycle as the target sources. Since these sources do not
vary on short (IDV) timescales, their measured amplitude variations can be
used to determine the time dependent gain variation of the whole system  
(telescope + receiver + atmosphere) with high ($<$ 0.8\,\%) accuracy. 
The observation of more than one of these secondary calibrators ensures the 
correctness of the assumption of non-variability of each secondary calibrator.

In the last step, the measured flux densities were then tied to the 
absolute flux-density scale, which was determined from repeated observations 
of the primary calibrators e.g. 3C\,48, 3C\,286, 3C\,295 and NGC\,7027 and using the flux-density scale of \cite{abs1} and \cite{abs2}. 
In Table \ref{tab:cal_flux} we summarize the used flux densities for each 
primary calibrator.

\begin{table}
\caption{\label{tab:cal_flux} Primary calibrators and their flux densities at the three observing frequencies.}
\centering
\begin{tabular}{cccc}
\hline 
\hline
Name & $S_{2.70\text{\,GHz}}$ [Jy] & $S_\text{4.85\text{\,GHz}}$ [Jy] & $S_{10.45\text{\,GHz}}$ [Jy]\\
\hline
3C\,48 & 9.32 & 5.48 & 2.60 \\
3C\,286 & 10.56 & 7.48 & 4.45 \\
3C\,295 & 12.19 & 6.56 & 2.62 \\
NGC\,7027 & & 5.48 & \\
\hline
\end{tabular}
\end{table}
 
The residual scatter in the measured flux densities of primary and 
secondary calibrator provides a very good and
also conservative estimate of the overall calibration accuracy. We note that
this is mainly limited by the systematic time variations of the 
overall system response, and not by the individual flux measurement, which is of
high SNR. These time dependent gain variations are monitored through the observed
variations of the primary and secondary calibrator fluxe densities. After correction
of all afore-mentioned gain effects, the residual variability of the 
calibrators yields a very reliable estimate of the measurement accuracy, and
furthermore does not depend on a-priori assumptions on receiver stability and 
telescope gain. We quantify the overall measurement accuracy
using the variability index $m_0= <(100\cdot\sigma/<S>)>$ of the secondary calibrators, 
where $m_0$ is the weighted average ratio of the standard deviations to the 
mean flux density for all calibrators. Thus, the uncertainty of an individual flux density
measurement is be given by $m_0 \cdot S$. The values for $m_0$ for each experiment
are given in Col. 5 of Table \ref{tab:idv_1128dat}. Typically, one obtains 
for Effelsberg
$m_0 \sim 0.4$\,\% at 2.70\,GHz, $m_0 \sim 0.5$\,\% at 4.85\,GHz, and $m_0 \sim 0.6$\,\% at 
10.45\,GHz, mainly reflecting the fact that the atmosphere limits the measurement
accuracy towards higher observing frequencies.

At Urumqi one obtains at 4.85\,GHz typically $m_0=0.5 - 0.7$\,\%. 
A somewhat larger value of $m_0 = 1.2$\,\% is observed in December 2005, which 
is due to bad weather conditions (rain and snow).

\subsection{Comparing the Effelsberg and the Urumqi data}

\begin{figure}
 \begin{center}
  \includegraphics[width=\columnwidth]{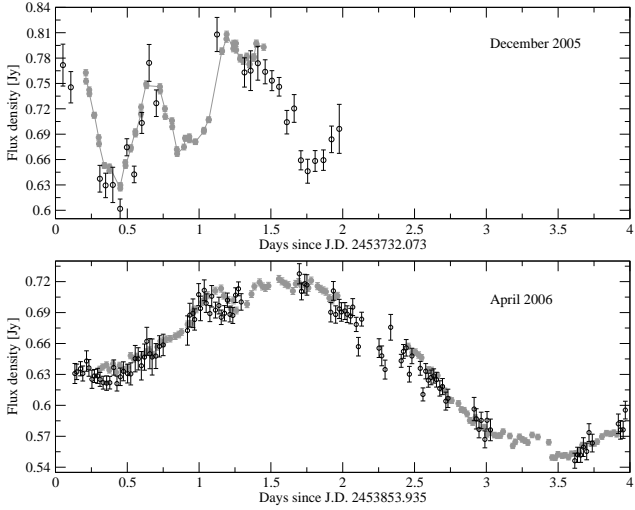}
  \caption{4.85\,GHz radio variability curves of J\,1128+5925
  obtained in December 2005 (top) and April 2006 (bottom). Filled grey circles 
  denote observations with the Effelsberg telescope, black open circles denote 
  observations carried out with the Urumqi telescope.}
\label{fig:EU}
 \end{center} 
\end{figure}

The direct comparison of the variability curves obtained
with the 100\,m Effelsberg and 25\,m Urumqi telescope in two quasi-simultaneous
observing sessions (Dec. 2005, April 2006,
see Fig. \ref{fig:EU}, and Tab. \ref{tab:idv_1128dat}),
shows a very good agreement of the variability curves measured at both
telescopes and demonstrates that the 25\,m Urumqi telescope is
well suited for IDV studies. The small differences seen in the rms-scatter of both
data trains are likely to result from some fundamental differences between
the two observatories: 
\begin{itemize}
\item The 100\,m Effelsberg dish uses the homology principle, 
the 25\,m Urumqi telescope does not. Consequently, the gravitational
deformation as a function of elevation is different for both dishes, resulting
in a lower aperture efficiency and steeper gain-elevation curves of the Urumqi
telescope. 
\item While for the 100\,m Effelsberg telescope, the focus position is automatically adjusted
at each elevation and also allows manual displacement and
refocusing at any time (day, night), the focus in the Urumqi telescope is 
at a fixed position, not allowing for additional focus corrections. This
leads to stronger variations of the overall time dependent gain. 
\item At 4.85\,GHz, Effelsberg uses a dual-horn beam-switch receiver, while at
Urumqi a single horn total power receiver is used. The Urumqi data are therefore
more affected by the weather than the Effelsberg data. At Urumqi,
this results in a less stable signal, which is probably the main reason for
the larger scatter seen in these data.
\end{itemize}

For the 6\,cm observations performed at Urumqi, we used the following
secondary calibrators, each of which with a flux density comparable
to that of our program sources:
B\,0836+710 ($S_{\rm 4.85\,GHz} = 2.1$\,Jy),
B\,0951+699 ($S_{\rm 4.85\,GHz} = 3.3$\,Jy),
B\,1203+645 ($S_{\rm 4.85\,GHz} = 1.2$\,Jy),
B\,1128+455 ($S_{\rm 4.85\,GHz} = 0.7$\,Jy),
B\,1311+678 ($S_{\rm 4.85\,GHz} = 0.9$\,Jy).
We find it remarkable that despite a 16 times smaller collecting area
of the Urumqi telescope, the achieved measurement accuracy ($m_0 = 0.5
-1.2$\,\% at Urumqi, $m_0=0.4-0.5$\,\% at Effelsberg, see Tab. \ref{tab:idv_1128dat}) is
only $\sim 2-3$ times worse than at Effelsberg. This demonstrates
that the IDV measurements performed with Urumqi are not limited by 
the size of the dish and the signal-to-noise
ratio, but by the larger fractional errors which are most likely 
the result of the afore-mentioned three effects.

\section{Data analysis} \label{analy}

For the variability analysis, we follow earlier work and use a parameterization
already introduced by \cite{idv_discovery2} and described in detail 
in e.g. \cite{reduc_idv2}.

To decide whether a source can be regarded as variable, 
we performed a $\chi^2-$ test, which tests the hypothesis 
whether the light curve can be modeled by a constant function. 
Sources for which the probability of this hypothesis is less than 
0.1\% are considered to be variable.

The amount of variability in a light-curve is described by the modulation 
index and the variability amplitude. The modulation index is defined 
by $m=100\cdot\frac{\sigma}{\langle \text{S} \rangle}$, where 
$\langle \text{S} \rangle$ is the time average of 
the measured flux density, and $\sigma$ is the standard 
deviation of the mean flux density. (The mean modulation index of the 
calibrators, $m_0$ describes the overall calibration accuracy, see Sect. \ref{obs}). 

In estimating the errors of the modulation index of the variable source, 
the statistical effects of a stochastic process have to be taken into account \citep{cimophd}. 
The errors originate from the limited observing time and from the time sampling.
If the observing time is much shorter than the
characteristic variability time-scale (no clear maximum followed by a
minimum can be observed in the light-curve), the variability index $m$ is
not well defined.
The contribution to the statistical error, originating from the 
length of the observing interval, is 
proportional to $\sqrt{t_\text{var}/T_\text{obs}}$, where $t_\text{var}$ is 
the characteristic variability time-scale
and $T_\text{obs}$ is the total observing time (B. Rickett, priv. comm.).
In addition, variability may also exist at times below the mean sampling 
interval of the measurements. Under-sampled variability also leads to
an underestimate of the variability index $m$. Assuming that significant 
variations may exist between three data-points, the error is given 
as: $3\cdot \langle t_\text{s}\rangle/T_\text{obs}$, where
$\langle t_\text{s} \rangle$ is the mean sampling interval \citep{cimophd}. 
Both of these effects were taken into 
account for the calculation of the uncertainty of the modulation index
$m$, which was found to range between $10$\,\% and $30$\,\% of the given 
value of $m$.

Following Heeschen et al. (1987) we define a $3\,\sigma$-variablity amplitude $Y$
through the noise-bias corrected $3\,\sigma$-value of the modulation index $m$. 
We define the variability amplitude of a given source with variability index
$m$ as $Y=3\sqrt{m^2-m_{0}^{2}}$, where 
$m_0$ is the mean modulation index of all secondary calibrators, 
which describes the statistical measurement accuracy during the observation.
$Y$ is set to zero for non-variable sources ($m=m_0$).

\subsection{Determination of the characteristic variability time-scales} \label{timescale}

There are many different definitions for the characteristic variability time-scale 
of IDV used in the literature. E.g. \cite{sf_def} use the structure function 
of the observed time series to define a characteristic variability time-scale. 
\cite{acf_def2} and \cite{acf_def} define the variability time-scale by the
half-width at half maximum of the autocorrelation function. 
\cite{decor_def2} use the half-width at $1/e$ of the maximum of the 
autocorrelation function \citep[this definition of the so called 
decorrelation time-scale is adopted from pulsar scintillation studies, 
see e.g.][]{decor_def}.
Since IDV light-curves appear often quasi-periodic, resembling some sort of 
sine-wave pattern, the variability time-scales can also be estimated from 
the mean peak-to-peak \citep{peak-to-peak} or peak-to-trough time \citep{0917annual2}.

To obtain reliable estimates of the variability time-scales present in our data, 
we derived the time-scales using 3 different methods: 

\begin{itemize}
\item We calculated from the light-curves the average peak-to-trough time ($t_\text{lc}$, Col. 12 
in Table \ref{tab:idv_1128dat}). The errors are calculated from the internal 
scatter of the different peak-to-trough times found in one data train. In 
some light curves, the peaks and troughs were not well-defined. In these cases 
we do not give an error estimate, the time-scales are given in parenthesis in the table.

\item We calculated the structure function (SF) for 
each time series and derived the ``saturation'' time-scale from it.
($t_\text{SF}$, Col. 13 in Table \ref{tab:idv_1128dat}). 
The SF is defined as 
$\text{SF}\left(\tau\right)=\langle \left(S\left(t \right)-S\left(t+\tau \right)\right)^2 \rangle$,
where $S\left(t\right)$ is the flux density time-series, $\tau$ is the time-lag and $\langle \rangle$
denote time averaging. Above the noise level, the SF rises monotonically
and is described by a power law. At large time-lags, the SF reaches its 
maximum at a ``saturation'' level, which is proportional to $2\cdot m^2$. 
The intersection of the power-law fit with this ``saturation'' level 
defines the characteristic variability time-scale $\tau$ \citep{beckert_evn}. 
The measurement error for this time-scale is derived from the formal errors of 
the power law fit and the fit of the ``saturation'' 
level \citep[][in prep.]{lars_thesis,lars_0716}.

\item We also calculated the autocorrelation function (ACF)
of the time series, using the method of \cite{corr_method}. The ACF is 
related to the SF via: 
$\text{SF}\left(\tau\right)=2\left(\text{ACF}\left(0\right)-\text{ACF}\left(\tau\right)\right)$.
Thus, the time-lag derived from the point where the SF reaches its 
maximum corresponds to the time-lag at the first minimum of the ACF. 
The derived time-scales ($t_\text{ACF}$) are given in Col. 14 of 
Table \ref{tab:idv_1128dat}. The error of $t_\text{ACF}$ are
determined from the scatter in the positions of the first minimum of the ACF,
when calculated with different bin sizes.
\end{itemize}

\begin{figure*}
 \begin{minipage}{0.5\textwidth}
 \begin{center}
  \includegraphics[width=9.7cm]{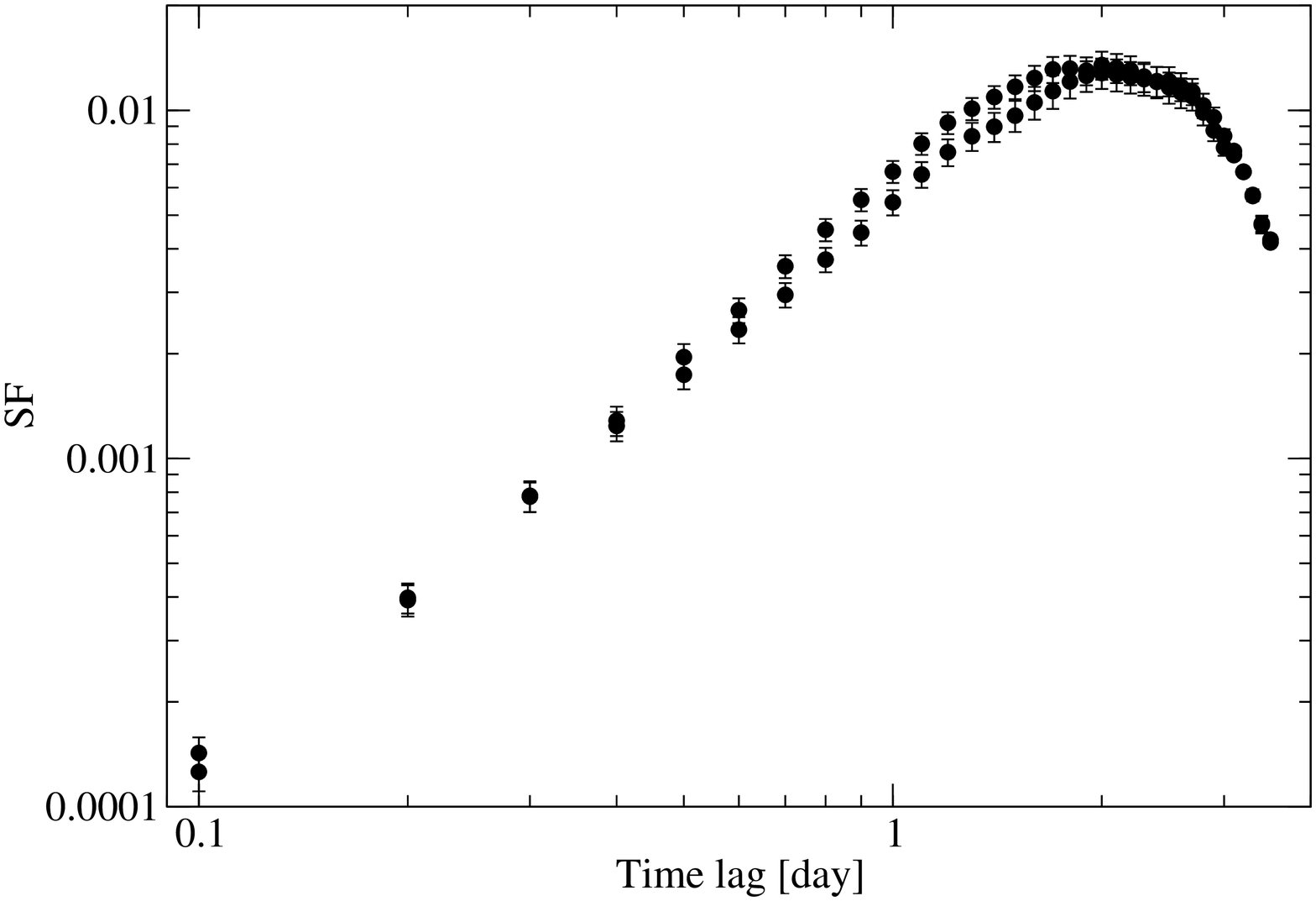}
 \end{center} 
 \end{minipage}
\hfill
 \begin{minipage}{0.5\textwidth}
\begin{center}
  \includegraphics[width=9.7cm]{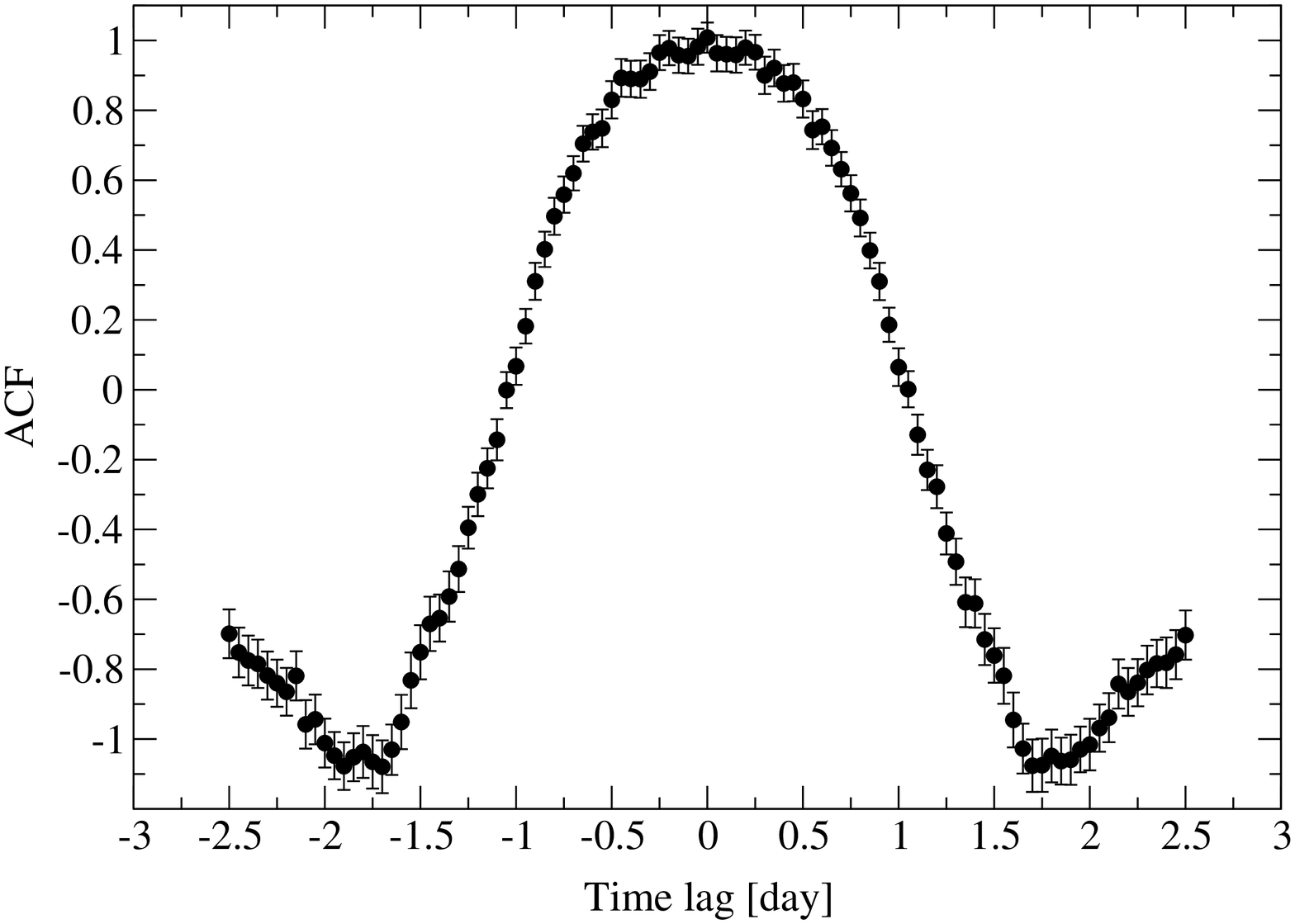}
\end{center}
 \end{minipage}
\caption{\label{fig:acf_sf}Example of a structure function (left) and an 
autocorrelation function (right) characterizing the variability of J\,1128+5925.
The data are from Effelsberg observations performed at 4.85\,GHz in April 2006.}
\end{figure*}

Examples of structure functions (SF) and autocorrelation functions (ACF) obtained 
for J\,1128+5925 are displayed in Fig. \ref{fig:acf_sf}. 
The functions were derived from the light curve observed at 4.85\,GHz with the 
Effelsberg radio telescope in 28 April 2006. The corresponding light-curve is shown 
in Fig. \ref{fig:lcs} in subplot (h). 


\section{Results} \label{res}

\begin{figure*}
 \begin{minipage}[t]{0.33\textwidth}
 \begin{center}
  \includegraphics[width=6.5cm]{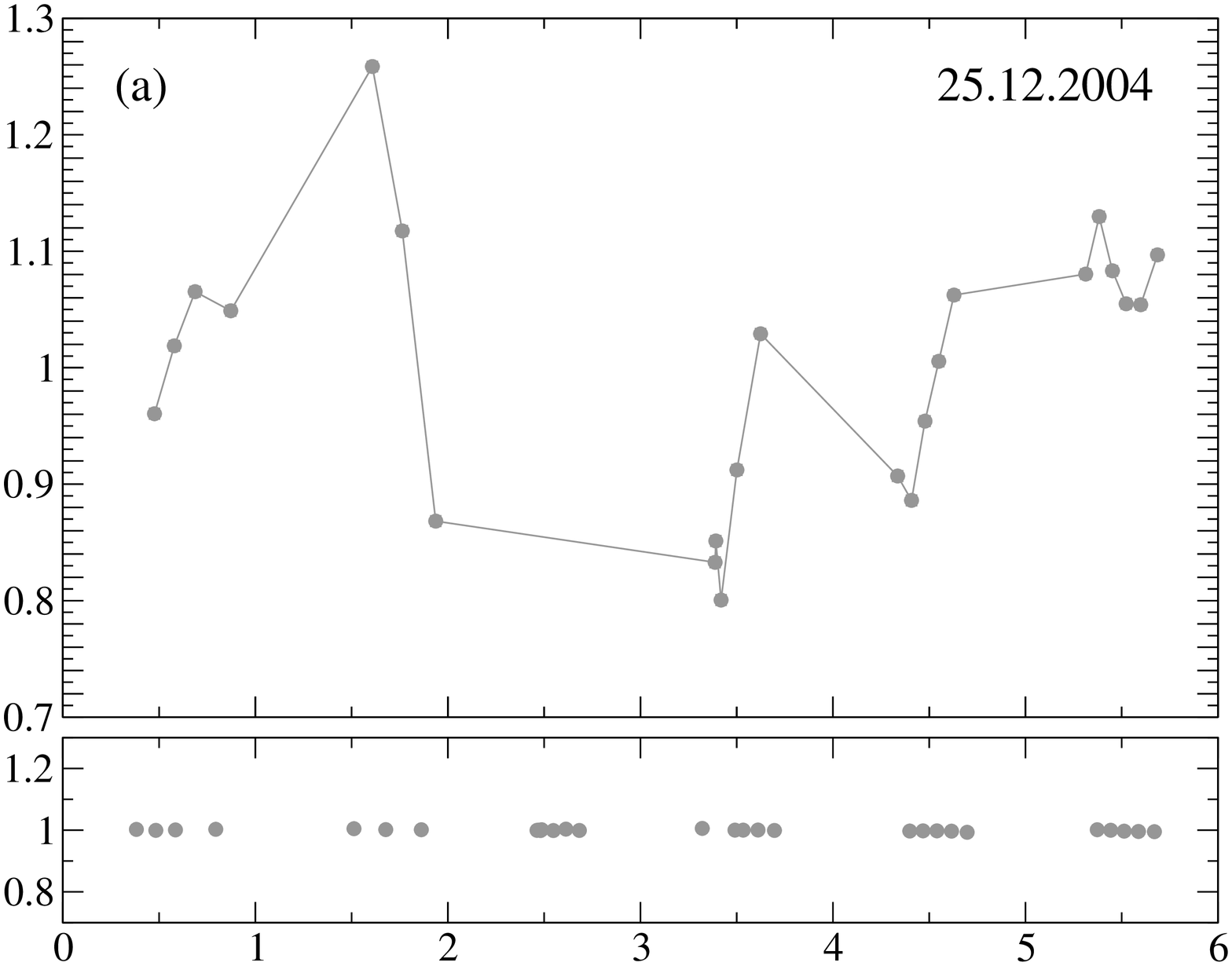}
 \end{center} 
 \end{minipage}
 \hfill
 \begin{minipage}[t]{0.33\textwidth}
 \begin{center}
  \includegraphics[width=6.5cm]{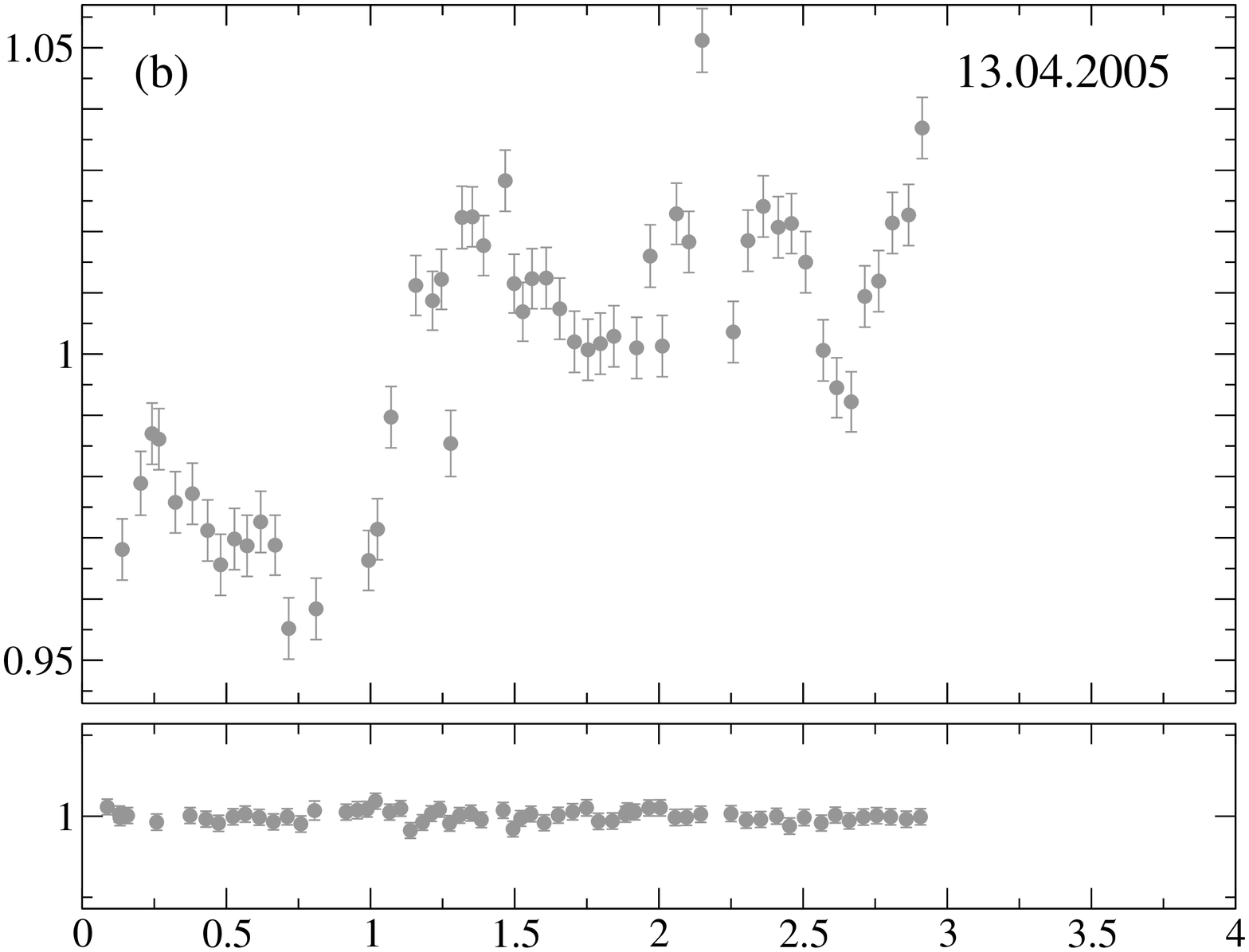}
 \end{center}
 \end{minipage}
\hfill
 \begin{minipage}[t]{0.33\textwidth}
 \begin{center}
  \includegraphics[width=6.5cm]{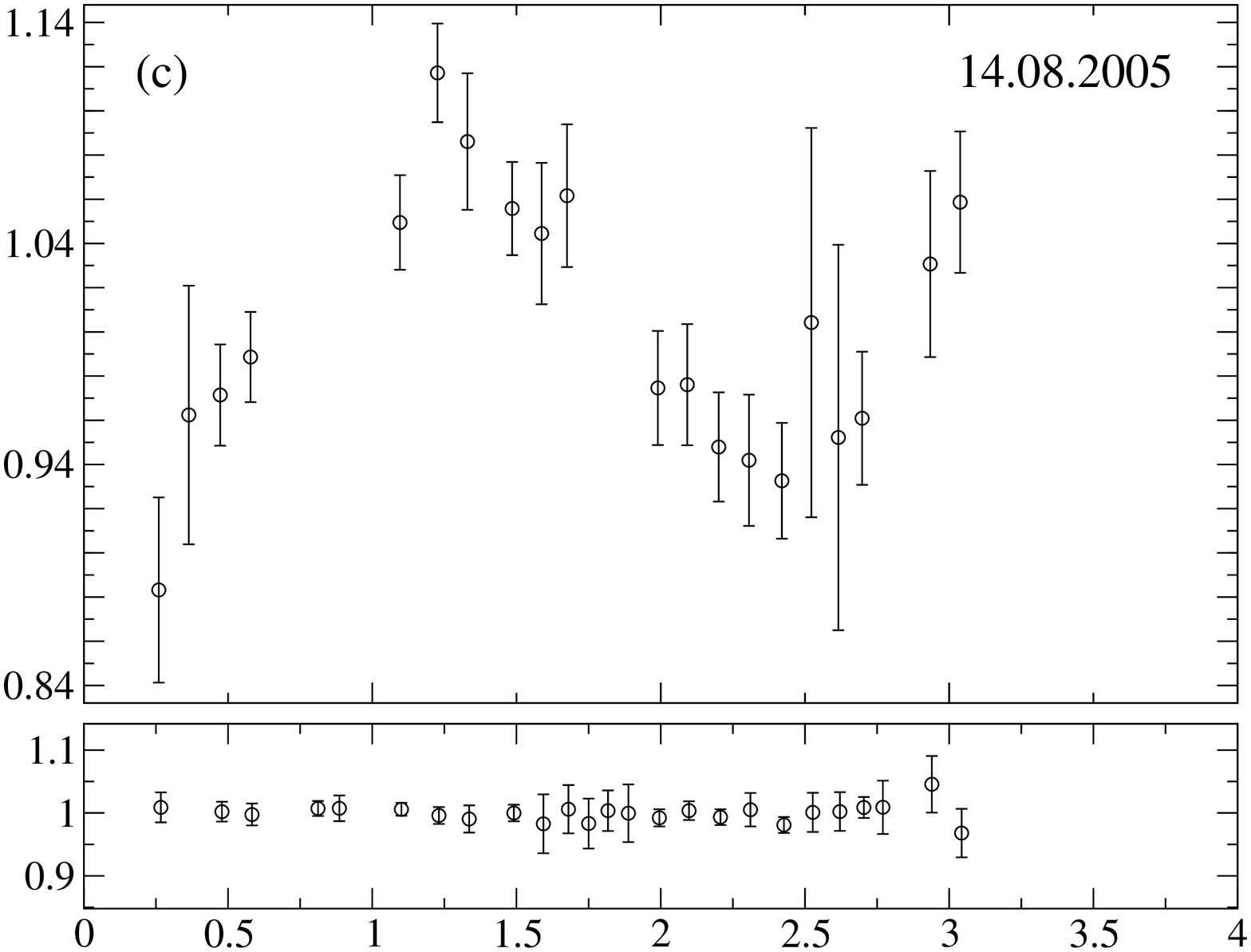}
 \end{center} 
 \end{minipage}
 \begin{minipage}{0.33\textwidth}
 \begin{center}
  \includegraphics[width=6.5cm]{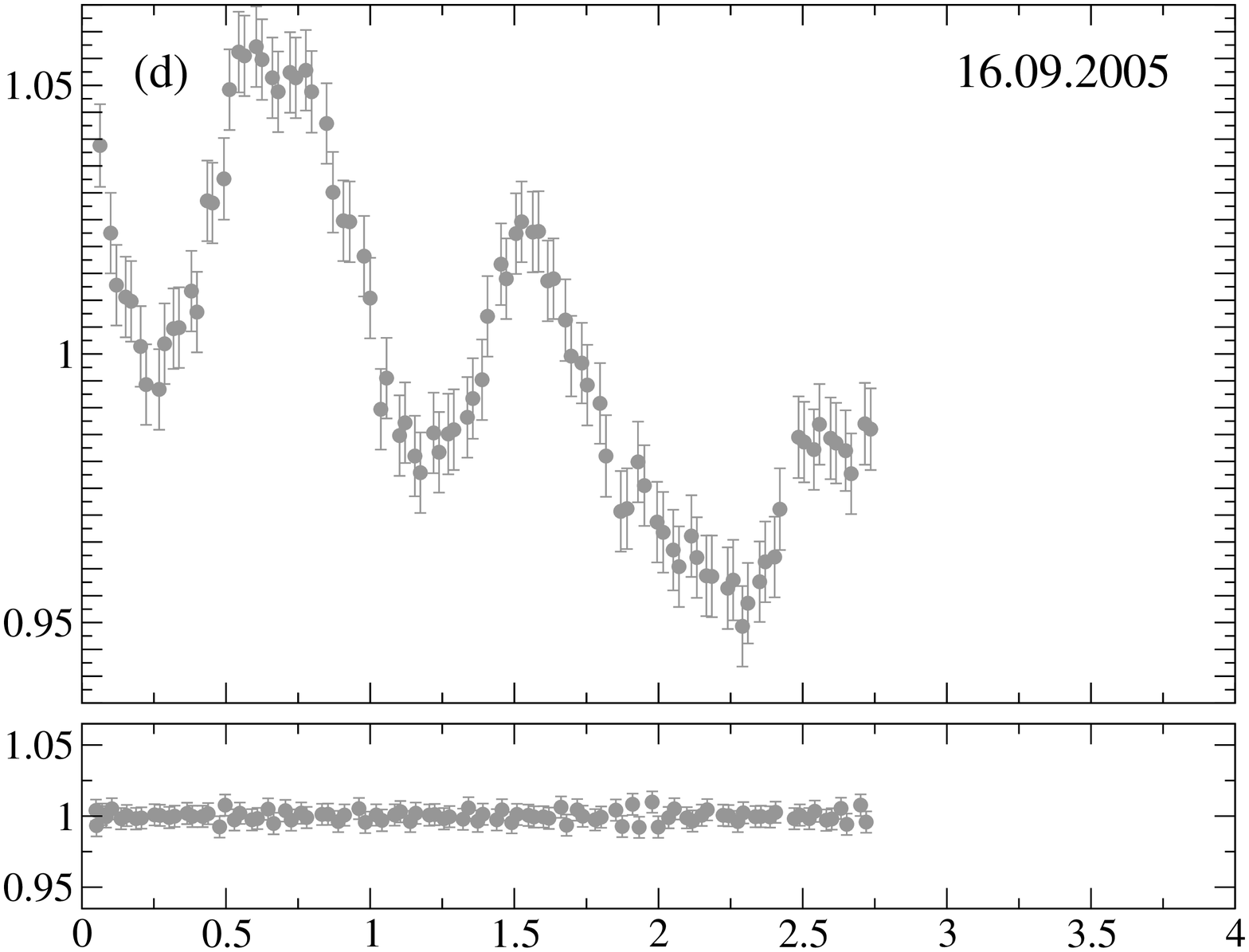}
 \end{center}
 \end{minipage}
 \begin{minipage}{0.33\textwidth}
 \begin{center}
  \includegraphics[width=6.5cm]{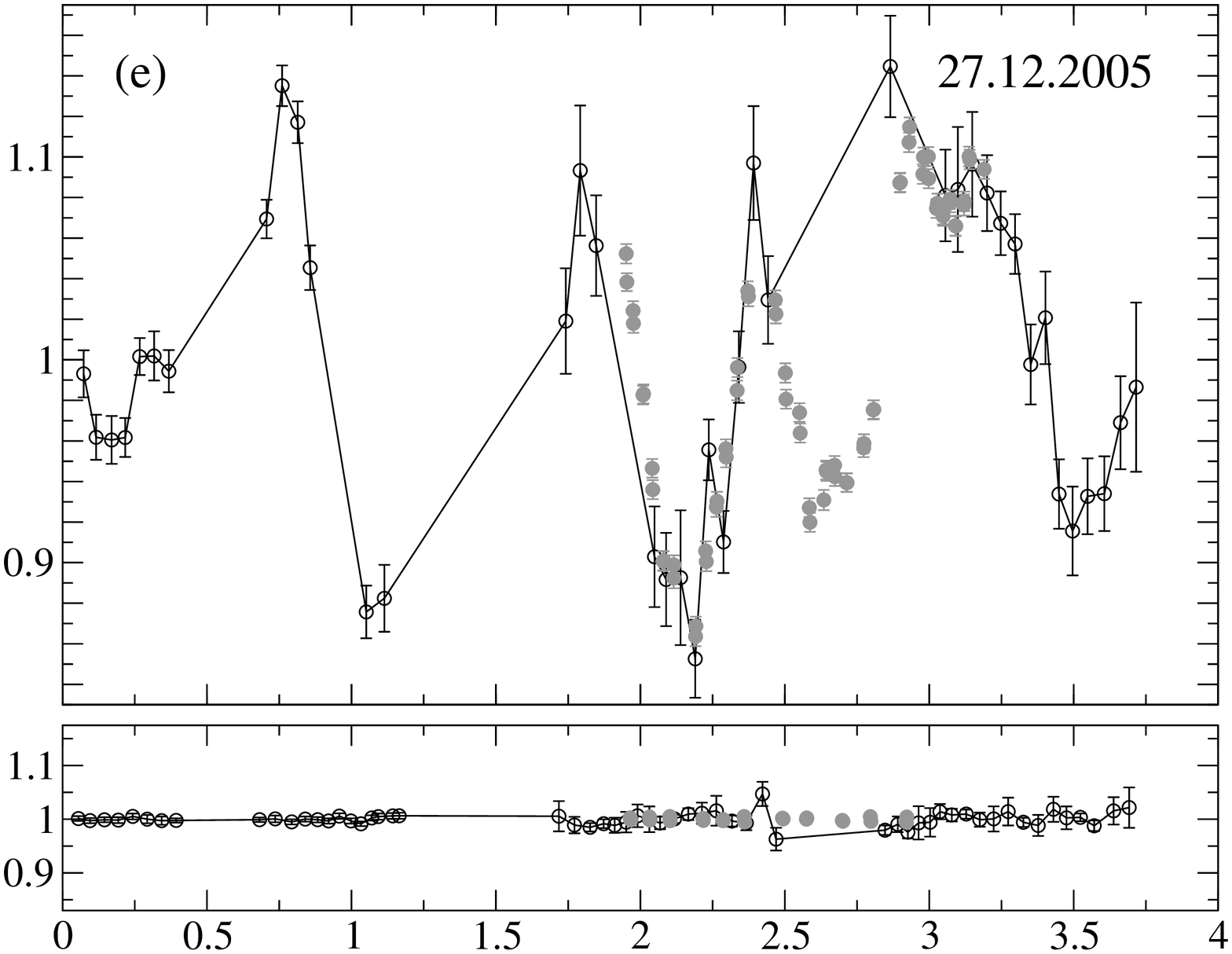}
 \end{center} 
 \end{minipage}
 \hfill
 \begin{minipage}{0.33\textwidth}
 \begin{center}
  \includegraphics[width=6.5cm]{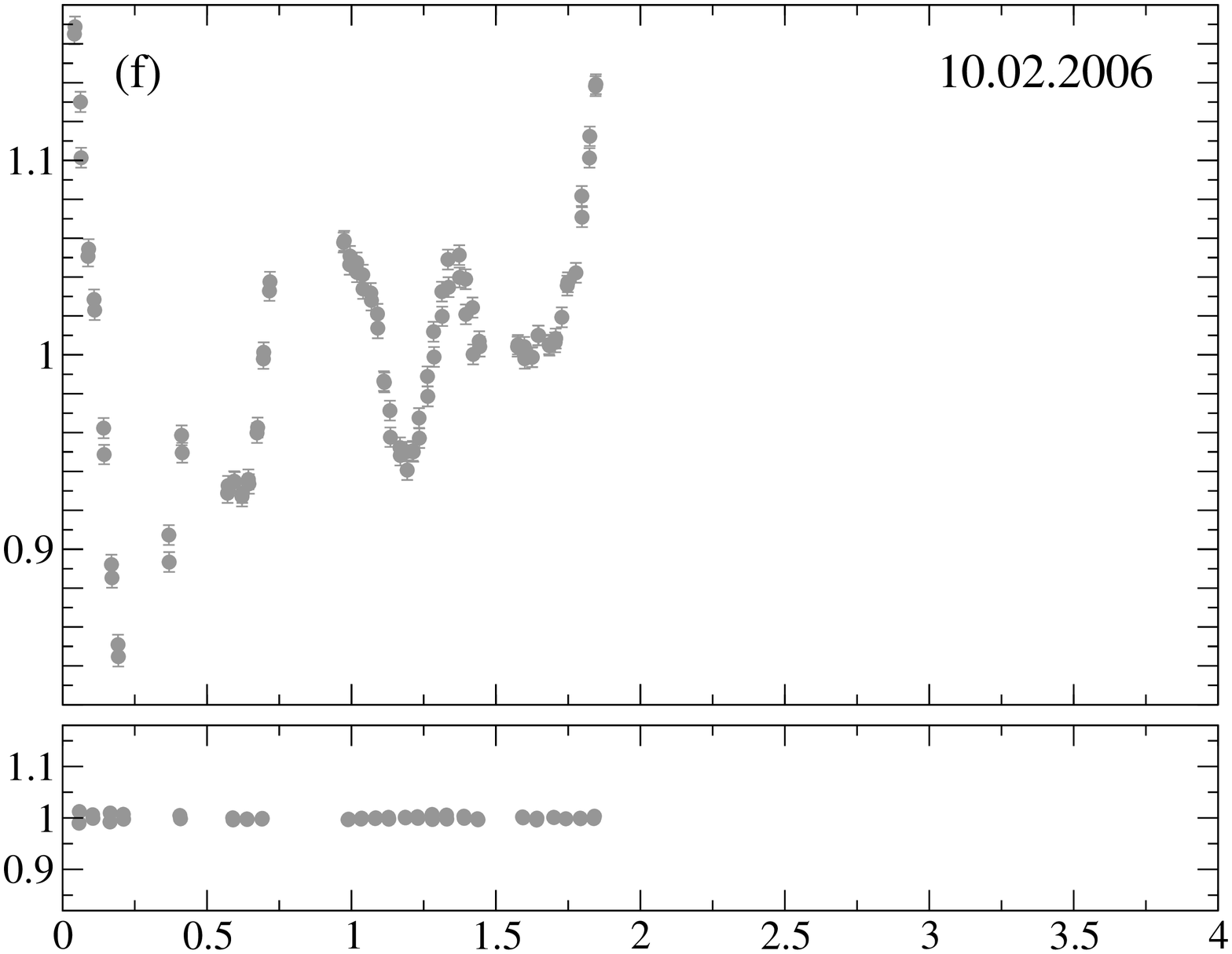}
 \end{center} 
 \end{minipage}
 \begin{minipage}{0.33\textwidth}
 \begin{center}
  \includegraphics[width=6.5cm]{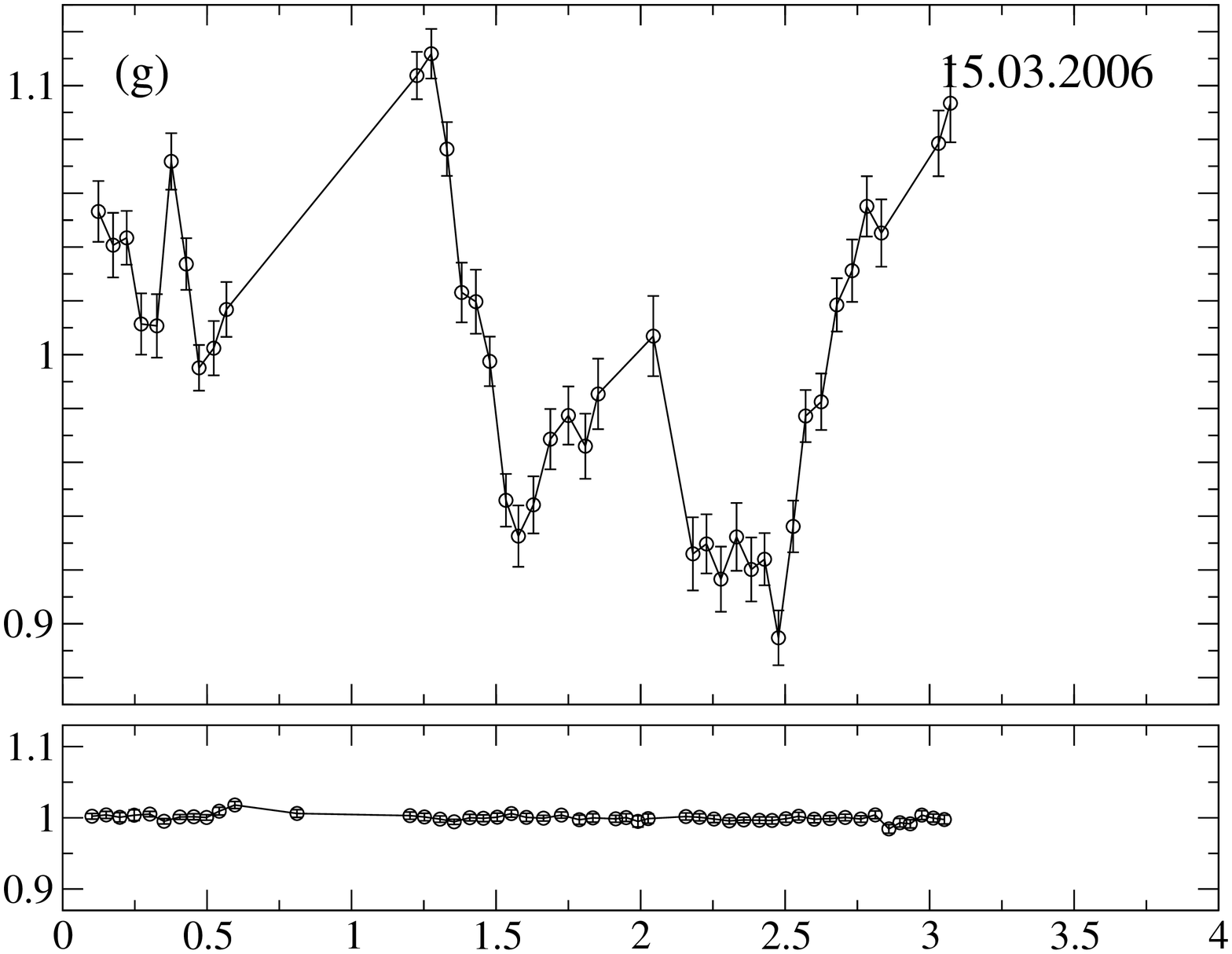}
 \end{center} 
 \end{minipage}
\hfill
 \begin{minipage}{0.33\textwidth}
 \begin{center}
  \includegraphics[width=6.5cm]{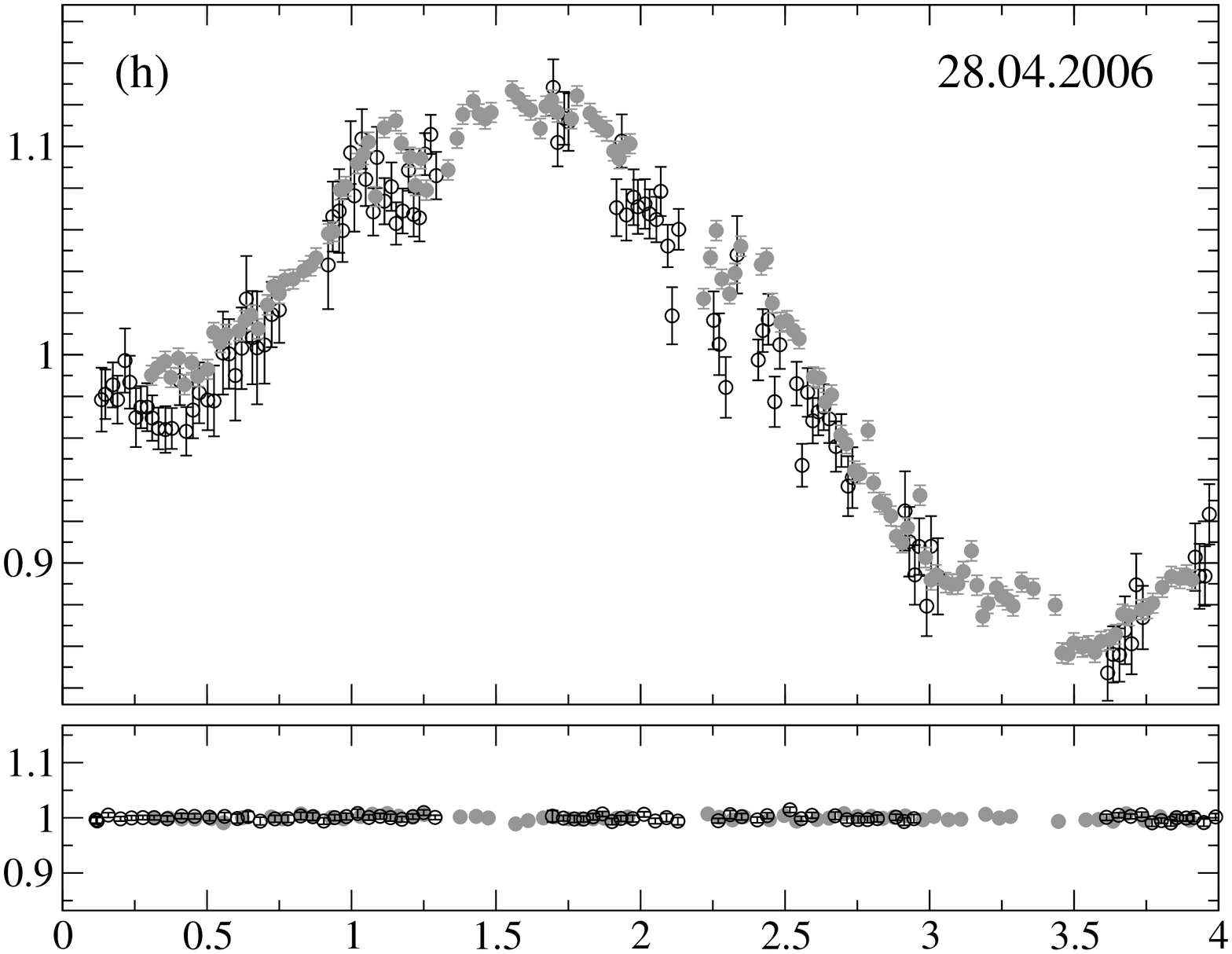}
 \end{center} 
 \end{minipage}
\hfill
 \begin{minipage}{0.33\textwidth}
 \begin{center}
  \includegraphics[width=6.5cm]{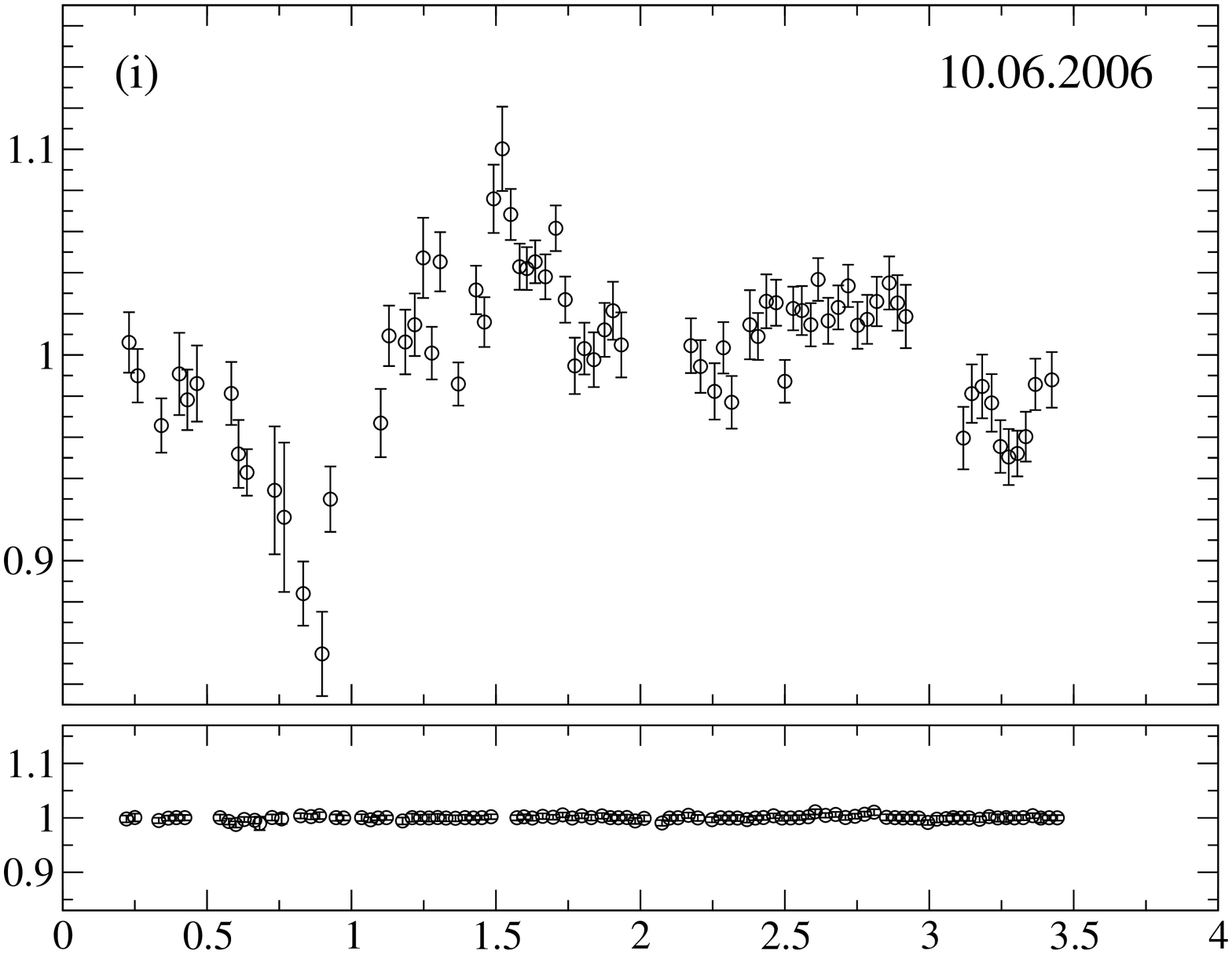}
 \end{center} 
 \end{minipage}
 \begin{minipage}{0.33\textwidth}
 \begin{center}
  \includegraphics[width=6.5cm]{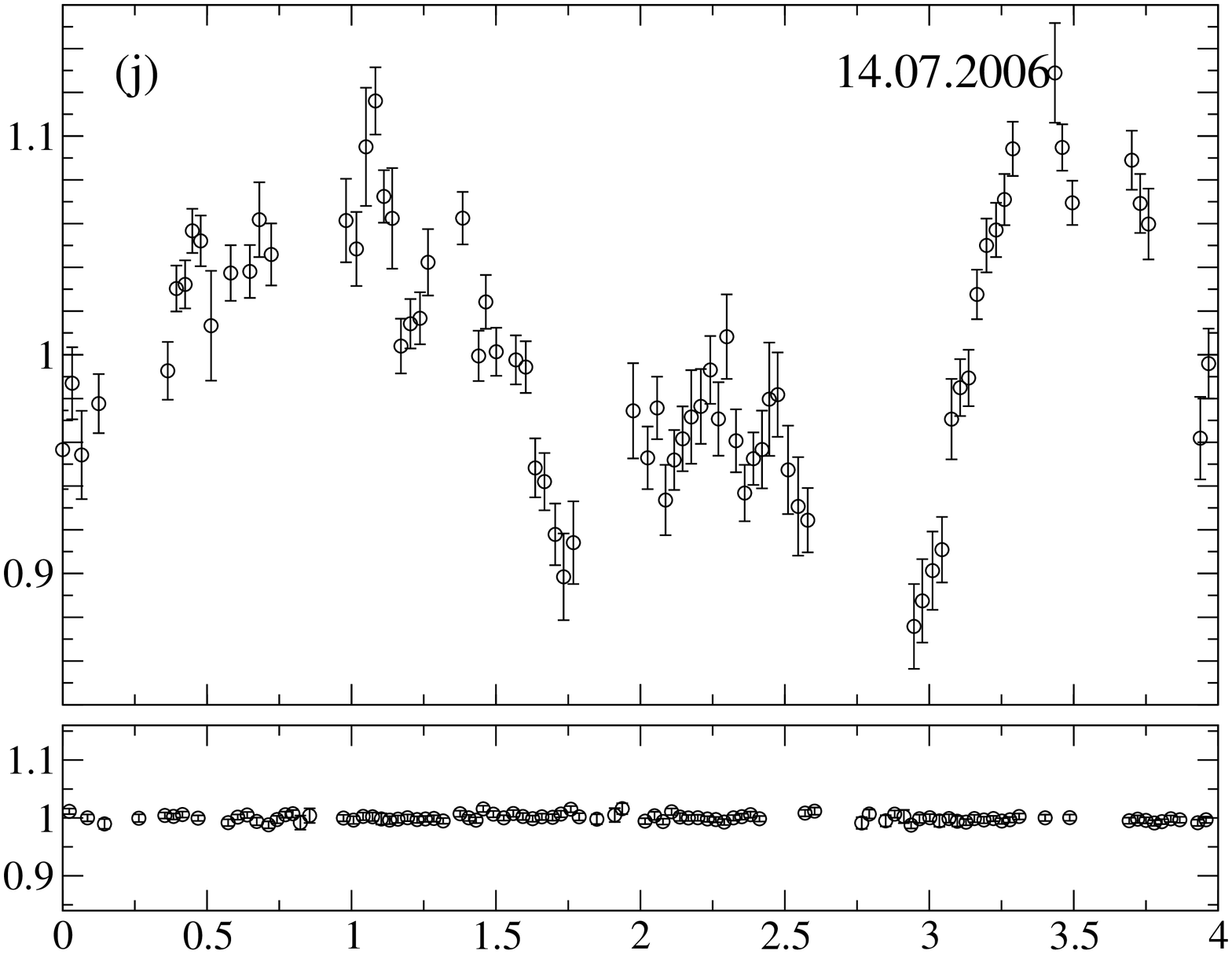}
 \end{center} 
 \end{minipage}
 \caption{4.85\,GHz light-curves of J\,1128+5925 (upper plots) and of 
 one secondary calibrator (B\,0836+710,lower plots) from December 2004 (panel a) until 
 July 2006 (panel j). Letters in brackets at the top left of each plot 
 refer to Table \ref{tab:obs1128}, 
 where the mean source flux density and the variability parameters are
 given. The abscissa displays the observing time [in days], relative to
 the starting date, which is shown in the top up right corner of each plot
 (see also Table \ref{tab:obs1128}). On the ordinate the normalized 
 flux densities are shown, with normalization by the mean source flux density. 
 Filled grey symbols denote Effelsberg observations, open black symbols 
 represent Urumqi observations.} 
 \label{fig:lcs}
\end{figure*}

In Figure \ref{fig:lcs} we show the radio variability curves of J\,1128+5925 obtained 
at 4.85\,GHz during December 2004 and July 2006. For each observing
date the normalized flux density of J\,1128+5925 (top panel) and of a
secondary calibrator (B\,0836+710, panel below) is plotted versus time (in days). In the plots,
the amplitude scale for J\,1128+5925 and for the secondary calibrators are similar,
visualizing the prominence of the variability seen in  J\,1128+5925.
Here, the peak-to-trough amplitude is found to reach $30$\,\%. 
The light-curves often appear quasi-periodic (e.g. in 
2005 September subplot (d), 2005 April subplot (h)) and show some complexity,
indicative of the presence of more than one variability time-scale. The variability 
characteristics, time-scales and strength of the variability change 
significantly from epoch to epoch, and sometimes even within a $3-4$\,day 
interval. For example in June 2006
(subplot (i)), where a pronounced trough and peak is followed by a much
shallower variations during the last days of the observations.


In Table \ref{tab:idv_1128dat}, we summarize the results of 
the observations of J\,1128+5925. We list the variablity parameters 
for all observing sessions:
Column 1 gives a label, which refers to the variability curve shown in
Figure \ref{fig:lcs}, col. 2 summarizes the observing date (first day, dd/mm/yyyy),
col. 3 the observing frequency, col. 4, the observatory, col. 5 the mean 
modulation index of the secondary calibrators, column 6 the number of data points 
obtained for J\,1128+5925, col. 7 the mean (time averaged) flux density of 
J\,1128+5925 in this session, col. 8 the standard deviation for the mean flux 
density, col. 9 the modulation index, col. 10 the variability amplitude, and
col. 11 the reduced $\chi^2$. For the degrees of freedom considered here,
a 99.9\,\% probability for variability is given, when the reduced $\chi^2$
exceeds $1.4-2.1$. With $\chi^2_\text{r}$ in the range of $\sim 4$ to $\sim 1000$,
the variability of J\,1128+5925 is highly significant
at all observing dates and frequencies. We note that at 10.45\,GHz the variations
are somewhat lower. A detailed description and discussion of the frequency 
dependence of variability is given in Sect. \ref{freq-dep}.

In columns 12, 13, and 14 of Table \ref{tab:idv_1128dat}, we summarize the 
three different estimates of the variability time-scale. The calculation 
of the variability time-scale is described in detail in Sect. \ref{timescale}. 

\begin{table*}
\caption{\label{tab:idv_1128dat}The variability parameters of J\,1128+5925 measured 
by the Effelsberg telescope and with the Urumqi telescope between 2004 and 2006. 
The letters in Col. 1 refer to the subplots of Fig. \ref{fig:lcs}. Col. 2 shows 
the starting date of the observation, Col. 3 the observing frequency, Col. 4 
the observing telescope ('E' for Effelsberg, 'U' for Urumqi). In Col. 5 the 
modulation index of the non-variable secondary calibrator is displayed, 
in Col. 6 the number of flux-density measurements for J\,1128+5925. 
The mean flux density, its standard deviation, the variability amplitude, 
and the modulation index of J\,1128+5925 are given in columns 7, 8, 9 and 10,
respectively. In Col. 11 the reduced $\chi^2$ ($\chi_\text{r}^2$) is shown. 
In Col. 12, 13 and 14 the measured variability times-scales are given. 
They were derived directly from the light-curves (Col. 12), 
using structure functions (Col. 13), 
and autocorrelation functions (Col.14). In some light-curves, 
the peaks and troughs are not well-defined. In these cases we do not give 
an error for $t_\text{lc}$ and the time-scales are given in 
parenthesis in Col. 12.}
\begin{minipage}{15cm}
\centering
\renewcommand{\thefootnote}{\thempfootnote}
\begin{tabular}{c|ccccc|ccccD{.}{.}{3}|ccc}
\hline
\hline
 & Starting date & $\nu$ & R.T. & $m_0$ & N & \(\langle S \rangle \) & \(\sigma \) & m & \multicolumn{1}{c}{$Y$} & \multicolumn{1}{c|}{\(\chi_\text{r}^2 \)} & $t_\text{lc}$ & $t_\text{SF}$ & $t_\text{ACF}$ \\
 & of observation & [GHz] & & [\%] & & [Jy] & [Jy] & [\%] & [\%] & & \multicolumn{1}{c}{[day]} & [day] & [day] \\
\hline
(a) & 25.12.2004 & 4.85 & E & 0.4 & 24 & 0.570 & 0.062 & 10.9 & 32.6 & 649.705 & $0.3 \pm 0.1$ & $\left(1.2_{-0.6}^{+0.4}\right)$ \footnote{The time-scale obtained from the SF analysis was not consistent with the light-curve.}& $0.3 \pm 0.1$ \\ 
(b) & 13.05.2005 & 4.85 & E & 0.5 & 54 & 0.611 & 0.013 & 2.2 & 6.4 & 19.932 & $\left(0.8\right)$ & $0.8^{+0.6}_{-0.5}$ &  $1.0 \pm 0.3$ \\ 
& 13.05.2005 & 10.45 & E & 0.6 & 29 & 0.730 & 0.028 & 3.8 & 11.2 & 11.875 & $\left(1.0\right)$ & $0.8 \pm 0.2$ & $1.0 \pm 0.1$ \\
(c) & 14.08.2005 & 4.85 & U & 0.6 & 20 & 0.682 & 0.040 & 5.9 & 17.5 & 4.291 & $1.2 \pm 0.2$ & $0.8 \pm 0.5$ & $1.1 \pm 0.2$ \\
(d) & 16.09.2005 & 4.85 & E & 0.5 & 91 & 0.719 & 0.021 & 2.9 & 8.6 & 14.815 & $0.5 \pm 0.2$ & $0.5 \pm 0.1$ & $0.6 \pm 0.15$ \\
(e) & 27.12.2005 & 4.85 & U & 1.2 & 40 & 0.713 & 0.056 & 7.9 & 23.4 & 25.872 & $ 0.3 \pm 0.2 $ & $0.4\pm0.2 $ & $0.34 \pm 0.04$ \\
(e) & 29.12.2005 & 4.85 & E & 0.4 & 64 & 0.725 & 0.052 & 7.2 & 21.5 & 237.791 & $0.3 \pm 0.1$ & $0.3 \pm 0.1$ & $0.30 \pm 0.06$ \\
 & 29.12.2005 & 2.70 & E & 0.3 & 31 & 0.524 & 0.036 & 6.8 & 20.4 & 57.033 & $0.6 \pm 0.2$  & $0.6_{-0.2}^{+0.3}$ & $0.54 \pm 0.02$ \\
 & 29.12.2005 & 10.45 & E & 1.4 & 46 & 0.826 & 0.027 & 3.2 & 8.7 & 5.340 & $0.2\pm0.1$ & $0.1^{+0.2}_{-0.1}$ & $0.3 \pm 0.1$ \\
(f) & 10.02.2006 & 4.85 & E & 0.4 & 96 & 0.723 & 0.049 & 6.8 & 20.3 & 192.638 & $0.2 \pm 0.1 $ & $0.1 \pm 0.05$ & $0.1 \pm 0.04$ \\
 & 10.02.2006 & 2.70 & E & 0.3 & 47 & 0.526 & 0.054 & 10.2 & 30.6 & 536.313 & $0.4 \pm 0.2 $ & $\left(1.1 \pm 0.5\right)$ \footnote{$t_\text{SF}$ are much longer than those derived from the ACF and the light-curve. There is an indication of an additional plateau in the SF at a time lag of $\approx 0.5$\,day. This agrees well with the other two values of the timescale.} & $0.5 \pm 0.1$ \\
(g) & 15.03.2006 & 4.85 & U & 0.5 & 40 & 0.668 & 0.038 & 5.7 & 16.9 & 28.835 & $0.5 \pm 0.1$ & $0.36\pm 0.18$ & $0.5 \pm 0.1$ \\
(h) & 28.04.2006 & 4.85 & U & 0.5 & 104 & 0.645 & 0.045 & 7.03 & 21.04 & 29.664 & $1.7 \pm 0.3$ & $1.3 \pm 0.2$ & $1.7 \pm 0.1$ \\
(h) & 28.04.2006 & 4.85 & E & 0.5 & 131 & 0.639 & 0.057 & 9.0 & 27.0 & 383.993 & $1.7 \pm 0.3$ & $1.4 \pm 0.1$ & $1.7 \pm 0.06$ \\
 & 28.04.2006 & 2.70 & E & 0.4 & 137 & 0.484 & 0.085 & 17.7 & 53.0 & 1107.124 & $1.9 \pm 0.2$ & $1.8 \pm 0.1$ & $1.8 \pm 0.2$ \\
(i) & 10.06.2006 & 4.85 & U & 0.5 & 72 & 0.595 & 0.024 & 4.1 & 12.2 & 7.925 & $\left(0.68\right)$ & $\left(0.5 \pm 0.3\right)$ & $\left(0.7\pm 0.08\right)$\footnote{The time-scales estimations are based upon the variations observed during the first one and half day of the observation. During the last two days the flux-density variations were much more reduced (see Fig. \ref{fig:lcs} (i) subplot).} \\
(j) & 14.07.2006 & 4.85 & U & 0.7 & 75 & 0.601 & 0.035 & 5.8 & 17.2 & 14.719 & $0.6 \pm 0.3$ & $0.6^{+0.3}_{-0.2}$ & $0.7 \pm 0.2$\\
\hline
\end{tabular}
\end{minipage}
\end{table*}

The typical variability time-scales range between $0.2 - 1.9$\,days. 
For each observing epoch and data set, the three different methods reveal
similar results and show the internal consistency of the measured variability 
time-scales.

Adopting a source-intrinsic interpretations for the variability, we can apply
the light travel-time argument and derive - via the source size - an apparent brightness 
temperature. Following \cite{light_size}, the angular size (in mas) is obtained 
from:
$\theta_\text{s}=3.56\cdot 10^{-4} t_\text{var} \left(1+z\right) \delta /D_\text{L}$. Here, $t_\text{var}$
is the variability time-scale (in days), $z$ is the source redshift, $D_\text{L}$  
the luminosity distance 
(in Gpc) and $\delta$ the Doppler factor. This size now can be used to
calculate a brightness temperature which, in the case of a Gaussian 
brightness distribution, is given by: $T_\text{B}=1.22 \cdot 10^{12} S \left(1+z\right)/\left(\nu^2\theta_\text{S}^2\right)$ $\text{K}$
\citep[e.g.][]{tb}. Here $S$ is the flux density in Jy, $\nu$ 
the observing frequency in GHz
and $\theta$ the source size in mas. 

Even if one uses the longest measured variability time-scale 
(of 1.7\,days as measured in April 2006 at 4.85\,GHz), one obtains
a brightness temperatures far in excess of the inverse-Compton limit 
of $T_\text{B} \sim 10^{19}$\,K. The fastest observed variations 
(seen in December 2005 and February 2006) imply $T_\text{B}\sim 10^{20}$\,K. 
Independent of the use of the inverse Compton brightness temperature 
limit \citep[$10^{12}$\,K, ][]{compton} or the equipartition brightness 
temperature limit \citep[$10^{11}$\,K, ][]{equipartition, equi2}, 
this leads to Doppler-boosting factors of a couple of hundreds to thousand, 
which are necessary to bring the observed brightness temperatures back to 
these limits.

\subsection{Time dependence of the variability}

In J\,1128+5925, the variability amplitudes and time-scales vary at each 
observing frequency significantly between the different epochs 
(see Fig. \ref{fig:lcs} and Table \ref{tab:idv_1128dat}). 

The length and sampling of the observations may influence the measurements 
of these parameters. However, only in some cases are the observed differences 
due to sampling effects. This is illustrated in the variability curves of
December 2005, which despite a different time sampling and and by a factor
of two different duration of the observation at Effelsberg and Urumqi still
leads to very comparable variability indices, amplitudes and time-scales.

\begin{figure}
 \begin{center}
  \includegraphics[width=\columnwidth]{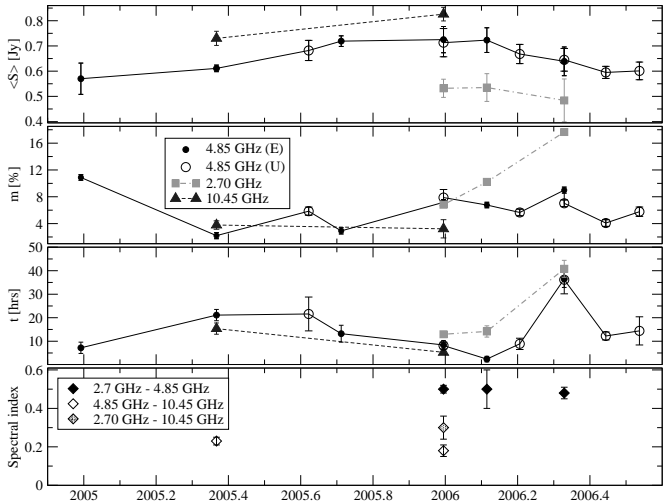}
  \caption{The mean flux density ($\langle S\rangle$ in Jy), 
  the modulation index ($m$ in percent), the characteristic variability 
  time-scale (in hours) and the spectral index of J\,1128+5925 
  displayed versus observing epoch. Black circles and black solid lines 
  denote observations at 4.85\,GHz: large open circles denote observations 
  with the Urumqi telescope, small filled circles denote observations 
  with the Effelsberg telescope. Triangles and dashed line denote observations 
  at 10.45\,GHz, grey squares with dashed-dotted line denote observations at 
  2.70\,GHz. Diamonds denote for the spectral index. Open diamonds are for 
  the spectral index between 10.45\,GHz and 4.85\,GHz, filled diamonds are 
  for the spectral index between 4.85\,GHz and 2.70\,GHz. 
  Grey diamond is for the spectral index between 10.45\,GHz and 2.70\,GHz.}
\label{fig:YST}
 \end{center} 
\end{figure}

To search for a possible correlation of the IDV with long-term variations of 
the total flux density, we plot in Fig. \ref{fig:YST} the mean flux density, the 
modulation index, the variability time-scale and the spectral index versus 
observing epoch. Measurements at different frequencies performed by different 
telescopes are shown with different symbols.

The variations of the modulation index and of the variability 
time-scale do not show a correlation with the
total flux density. The mean flux densities at 4.85\,GHz 
showed an increase by $20$\,\% until February 2006, over a
$1$\,yr time-scale. This increase is also seen at 10.45\,GHz. 
After this, the 4.85\,GHz flux density decreased almost to its original 
value by July 2006.
Unfortunately, we were unable to find additional total flux density monitoring data
for J\,1128+5925 in the literature, which would allow to us to check if and 
how the flux density varies on longer time-scales. Clearly, a future flux density
monitoring would be helpful.

The spectral index was calculated for those 4 epochs for which 
we had quasi-simultaneous multi-frequency measurements.
We obtained a spectral indices between 4.85\,GHz and 10.45\,GHz in May 2005
and in December 2005, and between 2.70\,GHz and 4.85\,GHz in February 2006 and April 2006. 
We also determined the spectral index using all three frequencies in December 2005. 
Within the limited amount of multi-frequency data, we do not see
pronounced temporal variations in any of these spectral indices. 
The observations, however, indicate some change of the spectral slope at 4.85\,GHz
from steep to flat. The data are also consistent with a spectral turnover
between 8\,GHz and 10\,GHz, which is already seen in Fig. \ref{fig:big_spektrum}.

\subsection{Annual modulation model} \label{sect_annmod}

The variability time-scales seen in
J\,1128+5925 are significantly different at different epochs 
(Fig. \ref{fig:lcs} and Table \ref{tab:idv_1128dat}).
At 4.85\,GHz, the changes in the characteristic 
variability time-scale do not correlate with other parameters of the variability (see Fig. \ref{fig:YST}). However, it is striking that in two observations separated by
one year (December 2004 and 2005) a very similar variability time-scale
of $0.3$ days is seen. The robustness of this short variability timescale is
independently confirmed by two telescopes (Effelsberg and Urumqi, December 2005).

\begin{figure*}
 \begin{center}
 \begin{minipage}[t]{8.1cm}
 \begin{center}
  \includegraphics[width=8.1cm]{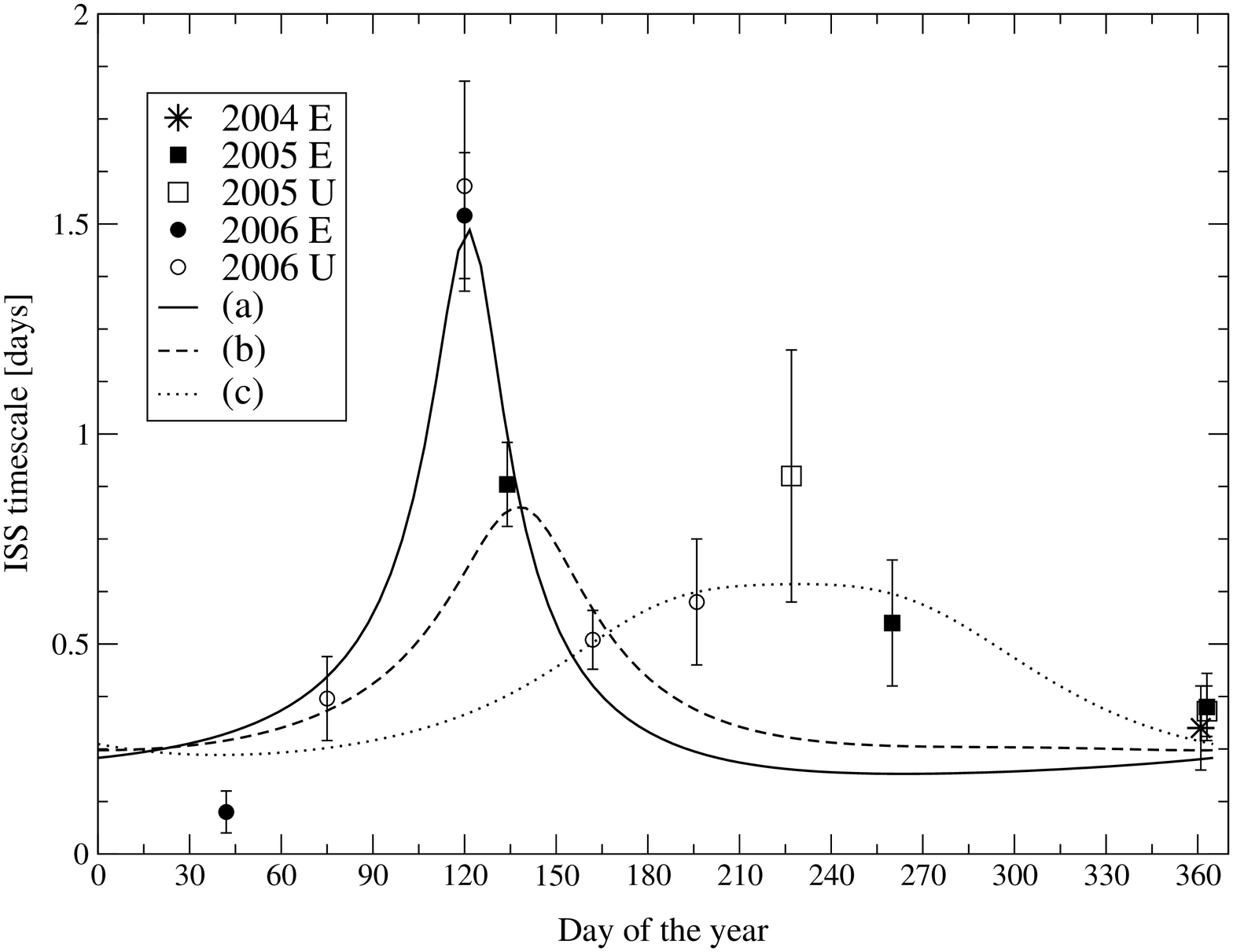}
 \end{center} 
 \end{minipage}
 \hfill
 \begin{minipage}[t]{8.1cm}
 \begin{center}
  \includegraphics[width=8.1cm]{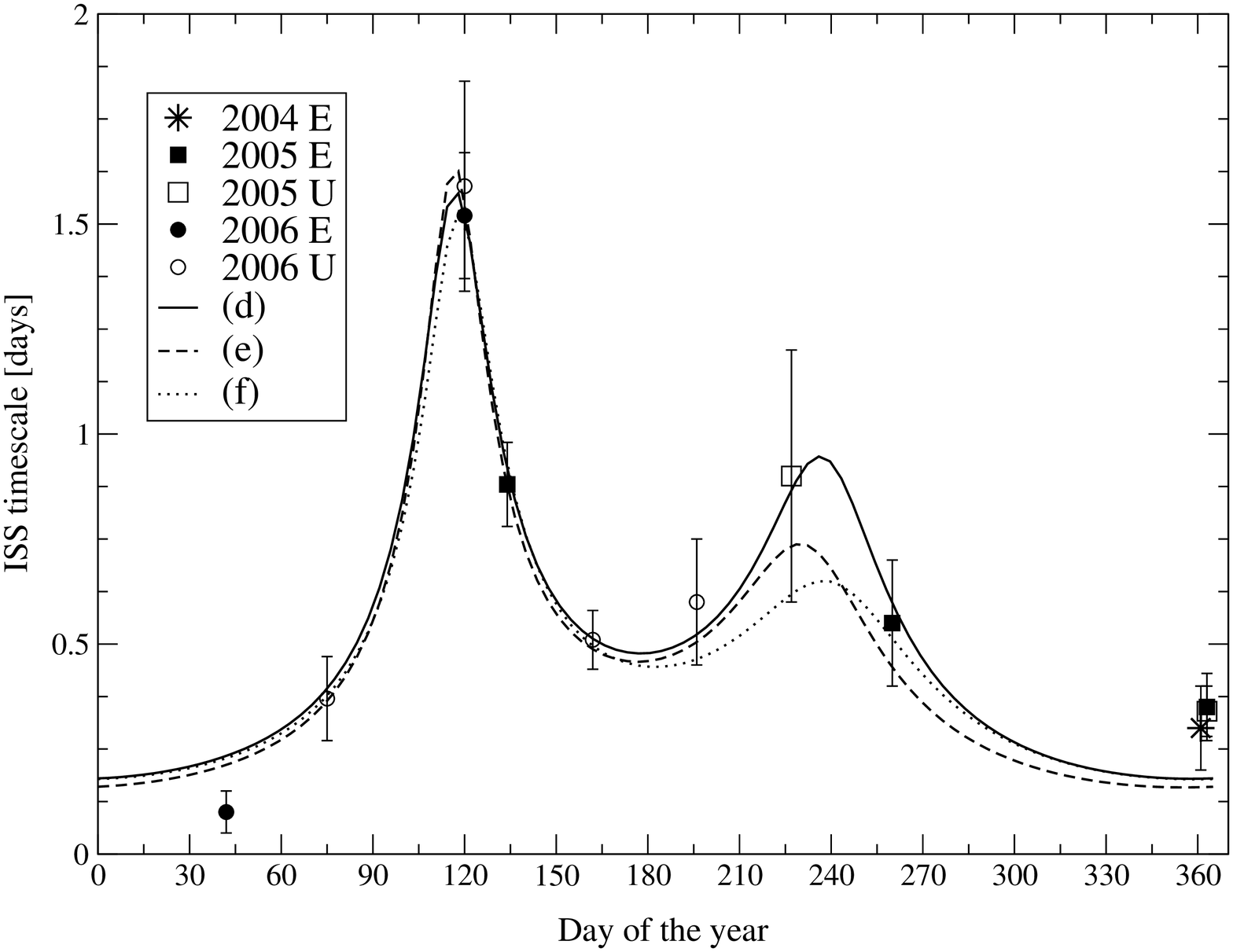}
 \end{center}
 \end{minipage}
 \caption{The IDV time-scale of J\,1128+5925 measured at 4.85\,GHz plotted 
 versus day of the year and fitted by 6 different annual modulation models. 
 Different symbols represent observations performed in different years: 
 stars stands for 2004, squares for 2005 and circles for 2006. Filled symbols 
 represent observations carried out with the Effelsberg telescope, open 
 symbols for observations with the Urumqi telescope. 
 Lines represent time-scales from an interstellar scintillation (ISS) model, assuming
 isotropy (left) and anisotropy (right). The labels refer to six different
 models, which are summarized in Table \ref{tab:fits}.}
\label{fig:annmod}
 \end{center} 
\end{figure*}

In Fig. \ref{fig:annmod}, we plot the variability time-scale versus day of 
the year. Different  symbols represent different years of observations: 
stars for 2004, squares for 2005, circles for 2006.
From Fig. \ref{fig:annmod}, the slowest variations occurs around day 120 
(end of April to early May, $t_\text{var}\approx 1.6$\,day). 
Although the number of data points is sparse, the figure shows another 
``slow-down'' of the variability time-scale around day 230 (August, September). 
In the following, we try to explain this behaviour by two scenarios within 
the framework of the annual modulation model. 

A seasonal cycle of the characteristic variability time-scale can 
be explained in a model of interstellar scintillation (ISS),
where the scattering material is regarded to be located in a thin plasma screen
at some distance from the Earth. The orbital motion of the Earth around the Sun
leads to annual velocity changes, resulting from the vector addition 
between the observer's velocity vector and the velocity vector of the screen. 
Thus annual modulation is a pure geometrical effect. A low relative velocity
between screen and observer results in a prolongation
of the observed variability time-scale. Half a year later the two velocity vectors
are oriented oppositely. This then leads to a large velocity difference 
and correspondingly to fast variability 
\citep[e.g.][]{bondi, 0917annual1, annual1819}.

In such a model the free parameters are the components
of the velocity vector of the screen in the plane perpendicular to the line of sight,
and the scattering length scale ($s_0$). The scintillation time-scale as a 
function of the day of the year ($T$) then is given as: 
$t\left(T\right) \sim s_0/\varv\left(T\right)$, where 
$\varv$ is the relative velocity between the observer and the screen.

In the left panel of Fig. \ref{fig:annmod} the three different lines represent 
annual modulation models calculated from three different sets of input parameters. 
The first curve, (a), is the best fit to the whole dataset and 
yields a $\chi^2_\text{r}=4.6$. 
In an attempt to fit the second ``slow-down'' of the variability time-scale 
around day 230, we first excluded the data points at day 120. This 
gave the slowest variability time-scale and thus determines the height of the first peak. 
Thus, we obtained curve (b). The $\chi^2_\text{r}$ of this fit was $4.1$, however including all the 
data points yielded $\chi^2_\text{r}=8.3$. 
To be able to fit the ``slow-down'' at day 230 one
has to exclude also the measurement at day 134. This is illustrated by curve (c)
The $\chi^2_\text{r}$ of this reduced dataset was $3.0$, however
using all the data points yielded a $\chi^2_\text{r}=13.3$.
To summarize, the first ``slow-down'' at day 120 is not due to one exceptionally long variability time-scale, but 
is constrained by two measurements, one obtained in May 2005 and the other in April 2006.

The parameters of the different curves are given in Table \ref{tab:fits}. 
In Col. 1, the letters refer to the corresponding curve 
displayed in Fig. \ref{fig:annmod}. In Col. 2 and 3, the components of the 
screen velocity are displayed
in Right Ascension ($\varv_\text{RA}$) and in Declination ($\varv_\delta$). 
In Col. 4, the scintillation length scale is given. 
Fig. \ref{fig:annmod} shows that this annual modulation
model can explain only one slow-down of the variability time-scale in one
year, but not two.

The previous model assumed an isotropic scintillation pattern. However, 
in a more general scenario, scintillation patterns could also be anisotropic. 
This is modelled by introducing some ellipticity. Including the anisotropy 
of the scintillation pattern introduces two more model parameters: the angular 
ratio of the anisotropy ($r$) and its position angle ($\gamma$). 
The scintillation time-scale now also depends on the direction of
the relative velocity vector (between the Earth and the screen) and 
how it ``cuts'' through the elliptical scintillation pattern.
Following \cite{bignall_newest},
the scintillation time-scale is given as the function of day of the year ($T$)
by the following expression:
\begin{equation}
t\left(T\right)=\frac{s_0 \sqrt{r}}{\sqrt{\varv\left(T\right)^2+\left(r^2-1\right)\left(\mathbf{v}\left(T\right) \times \mathbf{S}\right)^2}},
\end{equation}
where $s_0$ is the scintillation length scale.
$\mathbf{v}\left(T\right)$ is the relative velocity between the scintillation screen and the observer.
As the Earth orbits around the Sun $\mathbf{v}$ varies annually. $\varv$ denotes the absolute 
value of $\mathbf{v}$. $\mathbf{S}=(\cos{\gamma}, \sin{\gamma})$ is the unit vector, defining 
the orientation of the elliptical scintillation pattern, where the 
angle $\gamma$ is measured from East through North. 
The major and minor axis scale lengths of the scintillation pattern are defined
through $a_\text{maj}=s_0\sqrt{r}$, and $a_\text{min}=s_0/\sqrt{r}$, 
respectively. We note that in this paper the definition of the time-scale
differs from that used by \cite{bignall_newest}. This leads to a factor of $\sim 2.5$
difference between the scintillation length scale derived here and by 
\cite{bignall_newest}.

In the right panel of Fig. \ref{fig:annmod} the fits for the anisotropic 
annual models are displayed. The parameters of the three model curves are 
summarized in Table \ref{tab:fits}.

\begin{table*}
\caption{\label{tab:fits}Fit parameters for the annual modulation models shown 
in Fig. \ref{fig:annmod}. Col. 1 show labels used for the different fit curves. 
Col. 2 and 3 give the velocity components of the screen in Right Ascension and 
Declination direction. 
Col. 4 shows the scintillation length scale, Col. 5 the axial ratio, 
and Col. 6 the position angle of the scintillation pattern. 
In Col. 7 we give the reduced $\chi^2$ for the fits. 
For details of calculation of $\chi^2_\text{r}$ see text.} 
\centering
\begin{tabular}{c|ccccc|c}
\hline 
\hline
 & $\varv_\text{RA}$ & $\varv_\delta$ & $s_0$ & $r$ & $90^\circ-\gamma$ & $\chi^2_\text{r}$ \\
 & $\left[\text{km s}^{-1}\right]$ & $\left[\text{km s}^{-1}\right]$ & $10^5 \left[\text{km}\right]$ & & $\left[^\circ\right]$ & \\
\hline
(a) & $-15.0 \pm 1.7$ & $-9.7 \pm 1.7$ & $ 7.1 \pm 1.4$ & 1 (fixed) & 0 (fixed) & 4.6 \\
(b) & $-6.6 \pm 3.7 $ & $-10.3 \pm 3.3$ & $7.7 \pm 1.3$ & 1 (fixed) & 0 (fixed) & 4.1 (8.3) \\
(c) & $11.5 \pm 4.2$ & $-7.2 \pm 4.3$ & $9.9 \pm 1.3$ & 1 (fixed) & 0 (fixed) & 3.0 (13.3) \\
\hline
(d) & $-3.1 \pm 7.8 $ & $-11.5 \pm 1.9 $ & $ 11.8 \pm 2.5$ & $4.2 \pm 2.0$ & $-88.6 \pm 5.7$ & 3.0 \\
(e) & $-6.5 \pm 4.6 $ & $-12.0 \pm 1.1$ & $10.3 \pm 1.3$ & $4$ (fixed) &$-82.9 \pm5.7$ & 2.3 \\
(f) & $-6.5 \pm 4.6$ & $-11.3 \pm 2.1$ & $10.3 \pm 1.3$ & $3.3 \pm 1.2$ & $-88.6 \pm 11.5$ & 2.8 \\
\hline
\end{tabular}
\end{table*}

The simple isotropical scintillation model can not explain both,
the ``slow-down'' of the scintillation time-scale at around day 120, \emph{and} 
the increase of the variability  time-scales measured after day 230. 
The anisotropic scattering model of \cite{bignall_newest}, however, does 
reproduce both features quite well. Owing to its broader shape and lower
modulation, this second ``slow-down'' around day 230 is not as well determined 
as the peak around day 120, therefore are the corresponding 
parameters $r$ and $\varv_\text{RA}$ are not so well constrained.

In the case of a point source (i.e. a source smaller than the Fresnel angle) 
the scattering length scale is determined by the Fresnel
angle and the distance to the scattering screen. The Fresnel angle is given as $\theta_\text{F}=\sqrt{\lambda/(2 \pi D)}$, where
$D$ is the distance to the scattering screen and $\lambda$ is the observing frequency.
Extragalactic radio sources usually cannot be regarded as point-like
in this context, since $\theta_\text{s} > \theta_\text{F}$. Therefore,
the scattering length scale is mainly determined by the source size
and not by the Fresnel angle \citep[e.g.][]{goodman, walker}. Therefore,
the distance to the screen can be derived from the scintillation length scale
with the assumption of an angular source size.

For this, we use the VLBI data obtained for J\,1128+5925 through
images and Gaussian modelfits. The source was observed with VLBI at 5\,GHz
in the Caltech-Jodrell Bank Flat-Spectrum Survey 
\citep[CJF survey, ][ and S. Britzen in prep., 2007]{size1128,icrf}.
J\,1128+5925 is a very compact radio source, with a possible jet-like 
feature oriented towards the south-west at a position angle of $-100^\circ$. 
From circular Gaussian modelfits an upper limit to the VLBI core size of 
0.53\,mas (in 1992.731) and 0.13\,mas  (in 1994.702) (S. Britzen, priv. commm.)
is obtained.

An independent size estimate is obtained from the turn-over frequency of the 
radio spectrum of J\,1128+5925 (see Fig. \ref{fig:big_spektrum}). Assuming 
that the observed spectral turn-over is due to synchrotron self-absorption
and further assuming energy equipartition between particles and magnetic field,
allows to calculate the so called equipartition source size
\citep[e.g.][]{equipartition}. Adopting 8.35\,GHz as the turnover frequency 
and using the flux-density measured at this frequency ($0.539$\,Jy), 
one obtains the equipartition source size via: 
$\theta_\text{eq}=0.34 D_\text{L}^{-1/17} (1+z)^{9/17} \delta^{-7/17}$\,mas, 
where $D_\text{L}$ is the luminosity
distance, $z$ the redshift of the source and $\delta$ the Doppler factor. 
From the CJF survey (Britzen et al. 2007, and in prep.) the Doppler factor 
is estimated to be  $\delta \approx 5$. With this one obtains an 
equipartition size of $\approx 0.26$\,mas at 8.35\,GHz. We now may assume
that the source size is inversely proportional to the observing frequency 
(for discussion see, Sect. \ref{freq-dep}). This leads to an extrapolated
source size of $0.45$\,mas at 4.85\,GHz.

From the definition of the scintillation length scale $s_0=\theta_s D$, 
and using the above derived upper limit for $\theta_s$, we obtain a lower limit 
to the screen distance. Taking the source size derived from the spectrum of J\,1128+5925 
($0.45$\,mas), the calculated lower limits on the screen distance range between 
$10$\,pc and $20$\,pc.  
Using the source size derived from the modelfits of the CJF survey data ($0.13$\,mas) the lower limits on the screen
distances range between $\mathbf{36}$\,pc and $60$\,pc.
At these distances the Fresnel angular scales at 4.85\,GHz indeed range
between $15\,\mu$as and $ 36\,\mu$as. This confirms our initial assumption of
the Fresnel scale being smaller then the source size. Therefore the source 
cannot be regarded to be point-like.

If we assumed that the source size is smaller than or equal to the Fresnel scale, than
the scintillation length-scale is determined by the distance to the scattering 
screen and the Fresnel angular size. However, in that case the assumed source sizes
would be extremely small ($\sim 1\,\mu$as) compared to usual extragalactic source sizes.
Assuming a circularly symmetric source with FWHM corresponding to these Fresnel angle 
values implies a lower limit on the brightness temperature of the order of $10^{15}$\,K.

The modelfits of the CJF survey VLBI data revealed in two epochs (1994.702 
and in 1996.646) a secondary (jet?) component at a position angle $-95^\circ$ 
and $-104^\circ$ relative to the core. These position angles agree remarkably well
with the position angle of the anisotropy derived from the anisotropic annual 
modulation model ($\approx -90^\circ$). It is therefore possible that the
annual modulation and the by the model predicted amplitude and seasonal times
of the two major maxima in Figure \ref{fig:annmod} are related to the 
VLBI structure and its orientation. At the moment it is unclear, if the
agreement between both angles is a chance coincidence. Moreover the 
VLBI observations of the source were performed ten years before our IDV 
observations and it is unclear if the jet orientation is persistent over such
a long time. To confirm whether the orientation of the source elongation 
remained constant over time, and whether it is related to the anisotropy 
seen in the annual modulation pattern, requires further VLBI observations.

A seasonal cycles seen in the variability time-scales of some other 
IDV sources (such as J\,1819+3845 \citep{annual1819}, 
PKS\,1257-326 \citep{time_delay3}, 1519-273 \citep{annual1519_1}, 
B\,0917+624 \citep{0917annual1, 0917annual2} and S4\,0954+658 (L. Fuhrmann priv. com.)) 
are similarly explained in the framework of the annual modulation theory. 
Assuming anisotropy in the scintillation pattern, \cite{annual1819} 
and \cite{bignall_newest} successfully modeled the 
annual variations in the characteristic variability time-scales in the 
extremely fast IDV sources, 
J\,1819+3845 and in PKS\,1257-326. The derived screen distances in these sources 
were less than $\sim10$\,pc, comparable to the lower limits to the screen distance
derived here. 

\subsection{Frequency dependence of the variability} \label{freq-dep}

J\,1128+5925 has an inverted spectrum, with spectral index ranging between $\sim 0.2$ and $\sim 0.5$ 
at all epochs (Fig. \ref{fig:YST}).

Fig. \ref{fig:YST} also shows, for those epochs with simultaneous 
multi-frequency measurements, a systematic frequency dependence of
the modulation index, decreasing in magnitude from 2.70\,GHz to 10.45\,GHz.

\begin{figure}
 \begin{center}
  \includegraphics[width=\columnwidth]{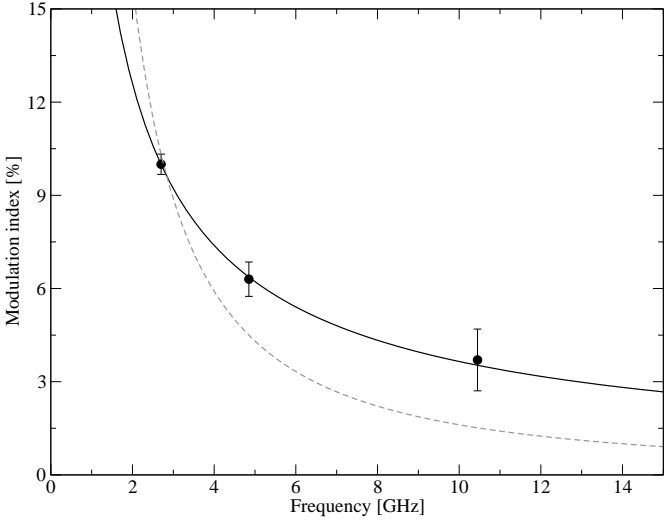}
  \caption{Frequency dependence of the average modulation indices. 
  The dashed grey line represents a fit to the data, where we assume a source 
  size smaller than or equal to the Fresnel scale. Thus $m \sim \nu^{-17/12}$. 
  The black, solid line shows a fit to a more realistic scenario, where we assume 
  that the source is extended and that its size changes with frequency following 
  a power law dependence.}
\label{fig:mod_freq}
 \end{center} 
\end{figure}

This frequency dependence is shown in Fig. \ref{fig:mod_freq}. It is
consistent with the ISS theory, which explains the phenomenon of IDV 
by interstellar scintillation in the so called weak regime 
\citep[e.g.][]{idv_wiss, GBI}. The variations attributed to refractive scintillation in 
the strong regime have longer time-scales, usually several days \citep[e.g.][]{idv_wiss,GBI}.
The characteristic time-scales of the variations  observed in J\,1128+5925 show a slight
increase with decreasing frequency in all of the multi-frequency datasets, 
which is also consistent with the theory of WISS (weak interstellar 
scintillation) (see Fig. \ref{fig:YST}). 

In the weak regime, $m$ decreases with increasing frequency. In the case of a 
point source (i.e.  $\theta_\text{s} \leq \theta_\text{F}$) \citep[e.g][]{goodman, walker}:
\begin{equation}
m=100\cdot \left( \frac{\nu_0}{\nu}\right)^{\frac{17}{12}}.
\end{equation}
where $\nu_0$ is the transition frequency between weak and strong scattering. 
We fitted this functional form to the three data points allowing $\nu_0$ to 
vary as a free parameter. The resulting fit is shown by the dashed line in Fig. \ref{fig:mod_freq}. 
The equation describing this fit is 
$m=(42.1 \pm 3.7) \left(\frac{1\text{\,GHz}}{\nu}\right)^{17/12}$.

If the source cannot be regarded as point-like, 
(i.e. $\theta_\text{s} > \theta_\text{F}$) the modulation index decreases with increasing source size
\citep[e.g][]{goodman, walker}:
\begin{equation}
m=100 \cdot\left(\frac{\nu_0}{\nu}\right)^{\frac{17}{12}} \left(\frac{\theta_\text{F}}{\theta_\text{s}}\right)^{\frac{7}{6}}. \label{extended_weak}
\end{equation}
Since the source size of most extragalactic sources is much larger than 
the Fresnel scale, the latter equation can be used
to describe the frequency dependence of the modulation index of IDV sources.
We also fitted this more realistic scenario, where we assumed that 
$\theta_\text{s}$ follows a power law dependence with the frequency. This fit is 
shown by the solid line in Fig. \ref{fig:mod_freq}. The equation describing this fit is
$m=(21.5 \pm 0.7) \cdot \left(\frac{\nu}{1\text{\,GHz}}\right)^{-0.77\pm 0.03}$.
Using equation \ref{extended_weak}, we can calculate the dependence of the source size on frequency: 
$\theta_\text{s} \sim \left(\frac{\nu}{1\text{\,GHz}}\right)^{-1.1\pm0.04}$. This is in good 
agreement, with the standard jet model \citep[e.g.][]{jet_model}. 
Here a constant brightness temperature is assumed for the self-absorbed jet base:
$T_\text{B} \sim S/(\nu^2 \theta_\text{s}^2) \sim \text{const}$. Thus, one obtains for a 
source with flat spectrum a source size, which is inversely proportional to the observing 
frequency.

Since the modulation index increases with decreasing frequency, 
we can give only an upper limit on the transition frequency between weak
and strong scattering of $\leq 2.7$0\,GHz.
From the annual modulation model we obtained the scintillation length scale, $s_0$, 
which is the product of the source size and the distance to the scattering screen. 
With known variability index $m$ from the fit and $s_0$, we now  
can calculate the distance of the scattering screen and the source size at 
4.85\,GHz. Since we have only an upper limit to the transition frequency, 
the resulting distance must be regarded as lower a limit, the source size as 
an upper limit. In Table \ref{tab:dist}, we summarize these values for the different 
annual modulation models. 
In every case, the source sizes are larger than the Fresnel angles, as expected 
for an extended extragalactic source and thus confirming the assumption that the scintillation 
is quenched by a finite source extent.

\begin{table}
\caption{\label{tab:dist} Lower limits to the screen distance (Col. 2) and upper 
limits to the source size (Col. 3). In the calculations we used the scattering 
length scale as derived from the annual modulation model, the frequency dependence 
of the source size, which was derived from the frequency dependence of the 
modulation index, and we assumed 2.70\,GHz as the upper limit to the transition 
frequency between weak and strong scattering. In Col. 1, the letters refer to 
the different annual modulation curves, see Table \ref{tab:fits} and 
Fig. \ref{fig:annmod}. In Col. 4, for comparison, we list the Fresnel angle 
at 4.85\,GHz at the distance of the screen. Col. 5 gives the estimated 
scattering measure for a transition frequency of $2.70$\,GHz.}
\centering
\begin{tabular}{c|D{.}{.}{1}ccc}
\hline 
\hline
 & \multicolumn{1}{c}{$D_\text{lower limit}$} & $\theta_\text{s}$ & $\theta_\text{F}$ & SM \\
 & \text{[pc]} & [mas] & [mas] & $10^{-5}\text{\,[m}^{-20/3}\,\text{kpc]}$ \\
\hline
(a) & $49.3$ & $0.10$ & $0.02$ & $7.7$ \\
(b) & $58.0$ & $0.09$ & $0.02$ & $6.7$ \\
(c) & $96.0$ & $0.07$ & $0.01$ & $4.4$ \\
\hline
(d) & $136.2$ & $0.06$ & $0.01$ & $3.3$ \\
(e), (f) & $103.8$ & $0.07$ & $0.01$ & $4.1$ \\
\hline
\end{tabular}
\end{table}

The estimates of the source size compare well with the core size measured
by VLBI (CJF survey, epoch 1994.702, $0.13$\,mas).
The equipartition source size is about a factor of $\sim 3$
larger than the ones obtained here. Doppler factors of 300 to 600 
would be needed to reduced these to the scintillating sizes. Another explanation is 
that the scintillating component is only the core, while the spectrum and the turnover frequency 
of J\,1128+5925 are influenced by the non-scintillating component as well. Therefore, the derived 
equipartition size might not represent the scintillating size.
 
According to \cite{tc93}, the transition frequency between weak and strong 
scattering is 
$\nu_0=185 \cdot \text{SM}^{6/17} \mathbf{(\frac{D}{1\text{\,kpc}})}^{5/17}$, 
where the Scattering Measure (SM) is defined by the line of sight integral 
of the amplitude of the electron density fluctuations ($C_N^2$) of
the turbulent ISM: $\text{SM}=\int_0^d C_N^2 dz$, 
where $d$ is the path length through the scattering medium.
Thus, knowing the distance, we can also calculate the SM.
In Table \ref{tab:dist} we show the SM derived in this way. 

\subsubsection{Cross-correlation between the frequencies}

\begin{figure*}
 \begin{minipage}[t]{0.5\textwidth}
 \begin{center}
  \includegraphics[width=8.1cm]{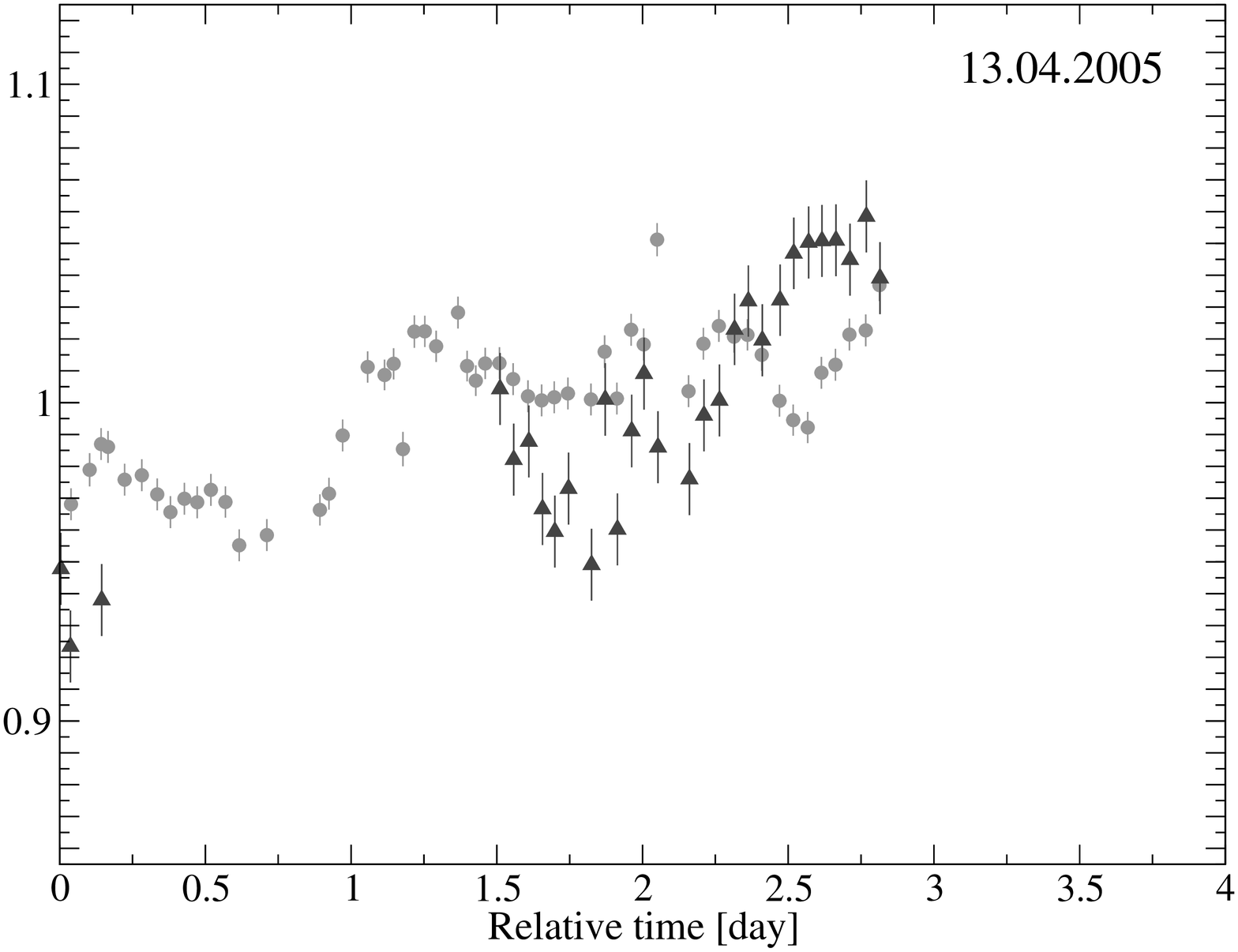}
 \end{center} 
 \end{minipage}
 \hfill
 \begin{minipage}[t]{0.5\textwidth}
 \begin{center}
  \includegraphics[width=8.1cm]{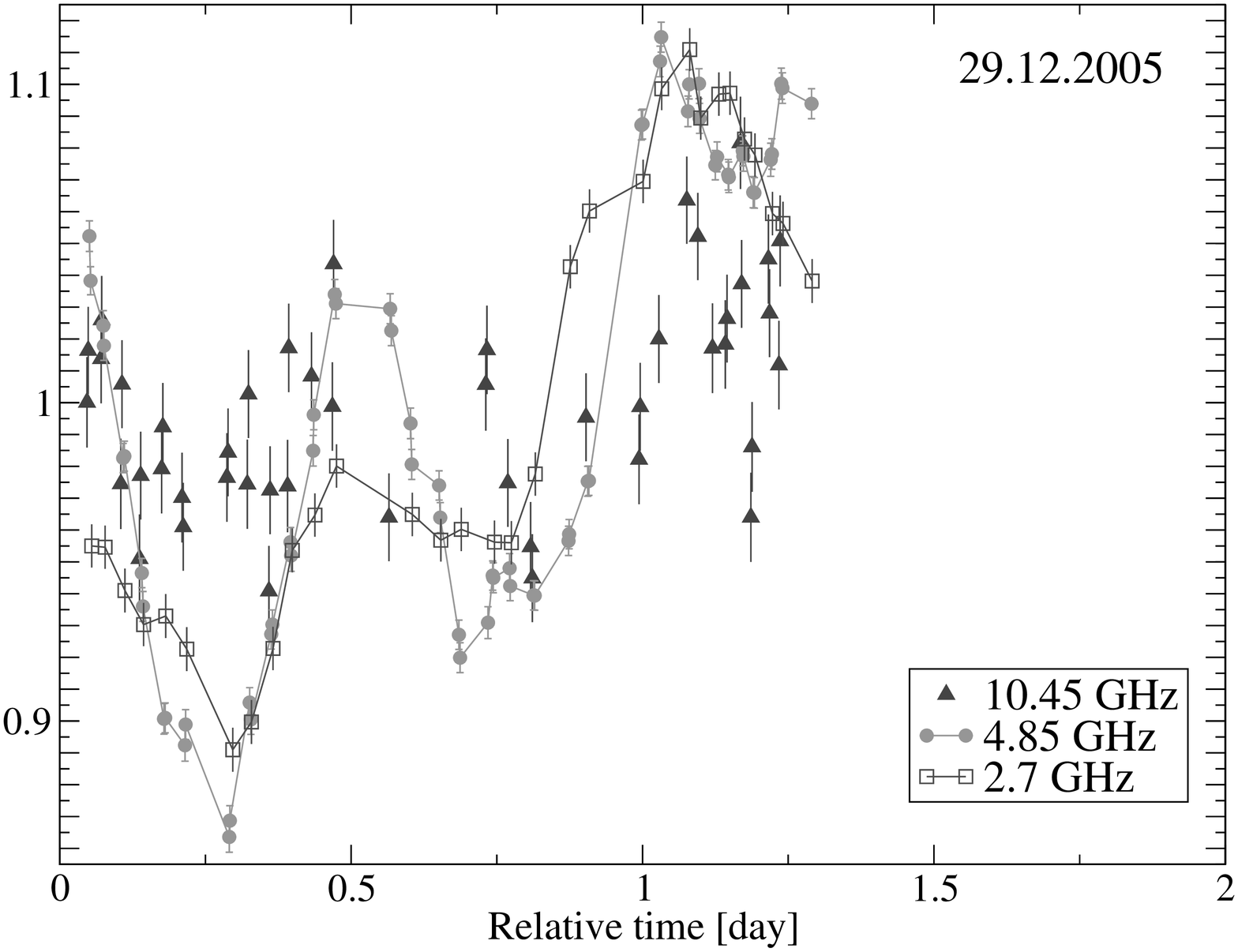}
 \end{center}
 \end{minipage}
 \begin{minipage}{0.5\textwidth}
 \begin{center}
   \includegraphics[width=8.1cm]{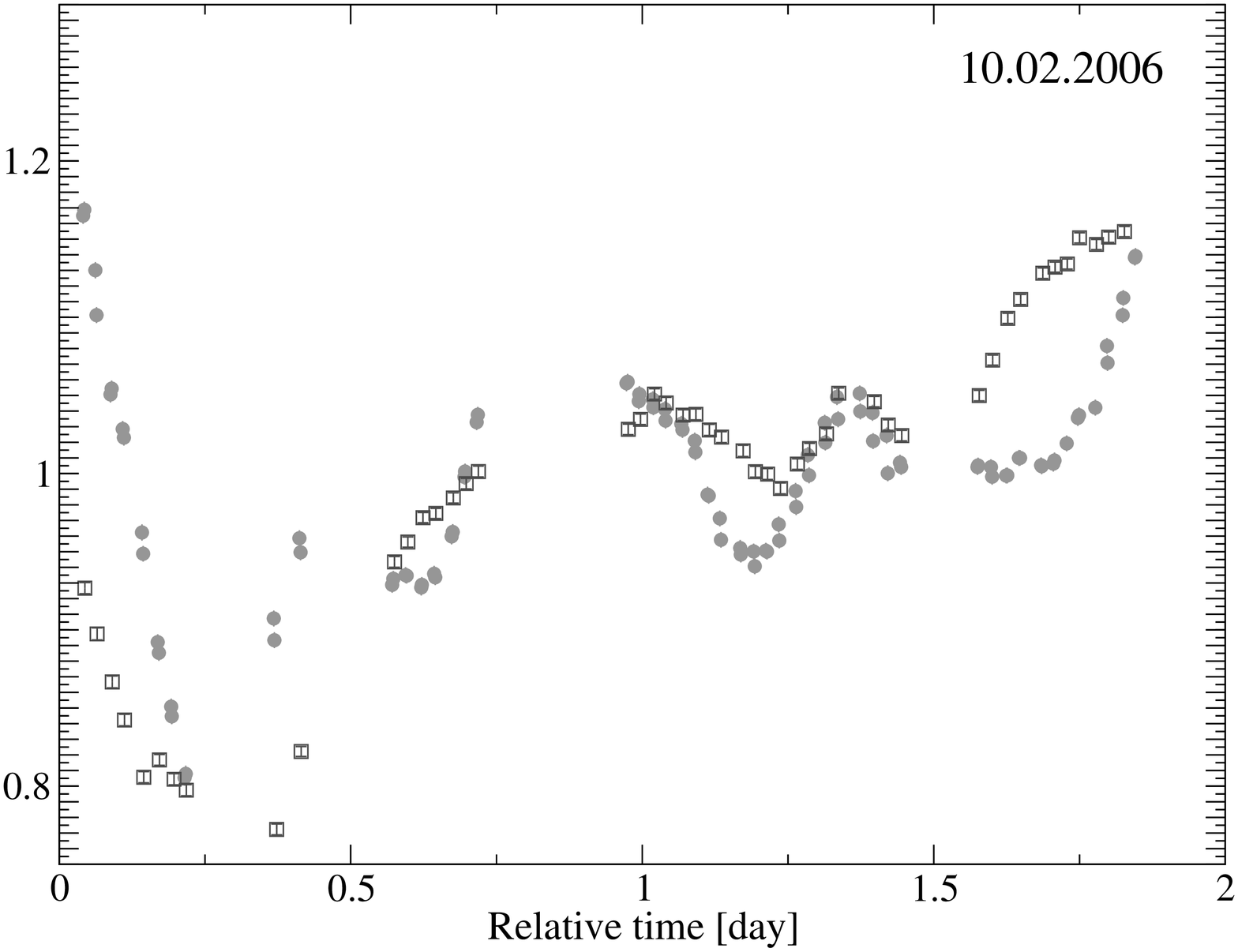}
 \end{center} 
 \end{minipage}
\hfill
 \begin{minipage}{0.5\textwidth}
 \begin{center}
  \includegraphics[width=8.1cm]{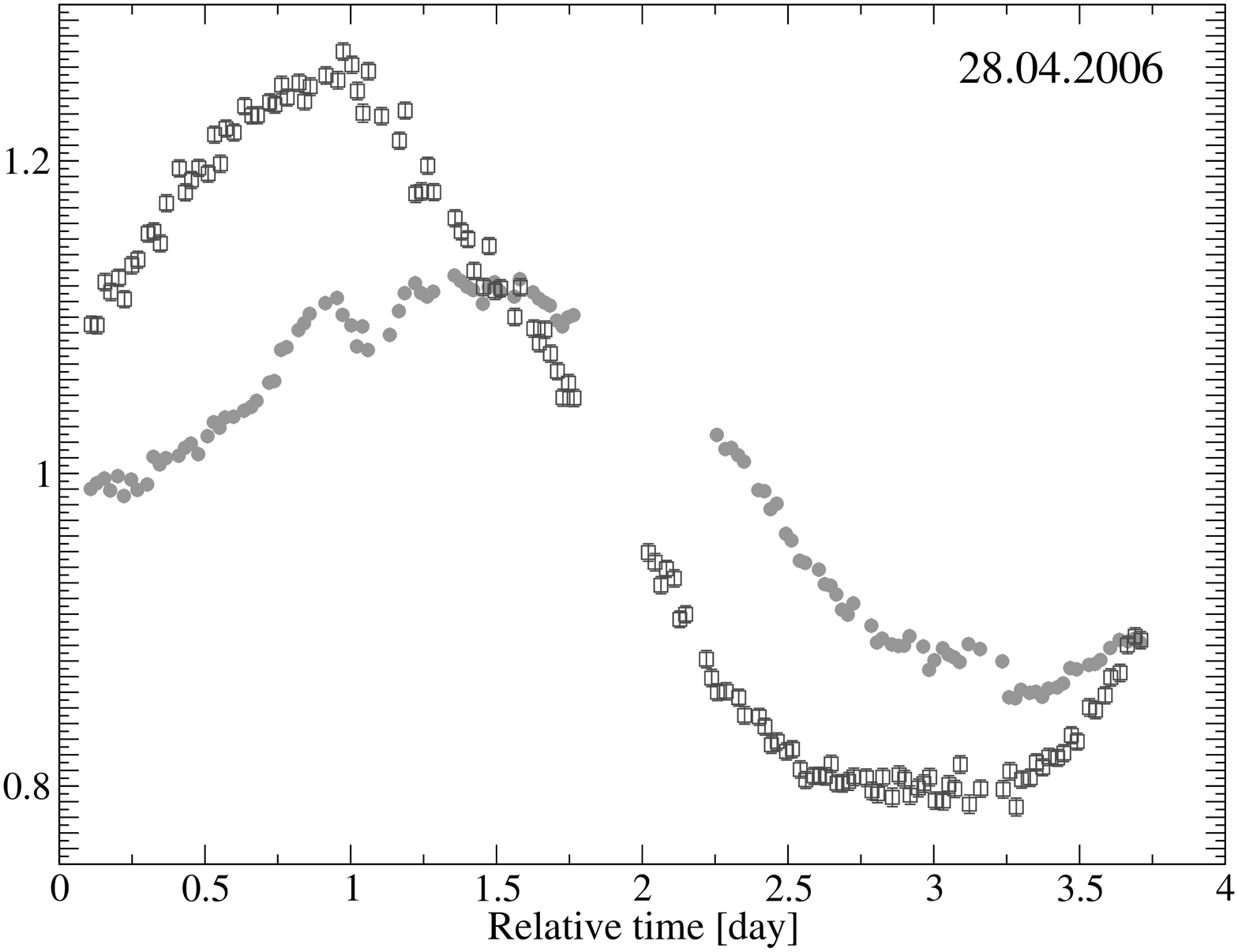}
 \end{center}
 \end{minipage}
 \caption{Light-curves of J\,1128+5925 at those dates when measurements were carried 
 out at more than one frequency quasi-simultaneously. On the abscissa the 
 observing time in days is displayed, the starting dates are given in the 
 upper right corner of each plot. On the ordinate the normalized flux densities 
 are given. Filled grey circles represent the 4.85\,GHz measurements, open 
 black squares represent the 2.70\,GHz measurements, black filled 
 triangles represent 10.45\,GHz measurements.}
\label{fig:freq}
\end{figure*}

\begin{figure}
 \begin{center}
  \includegraphics[width=\columnwidth]{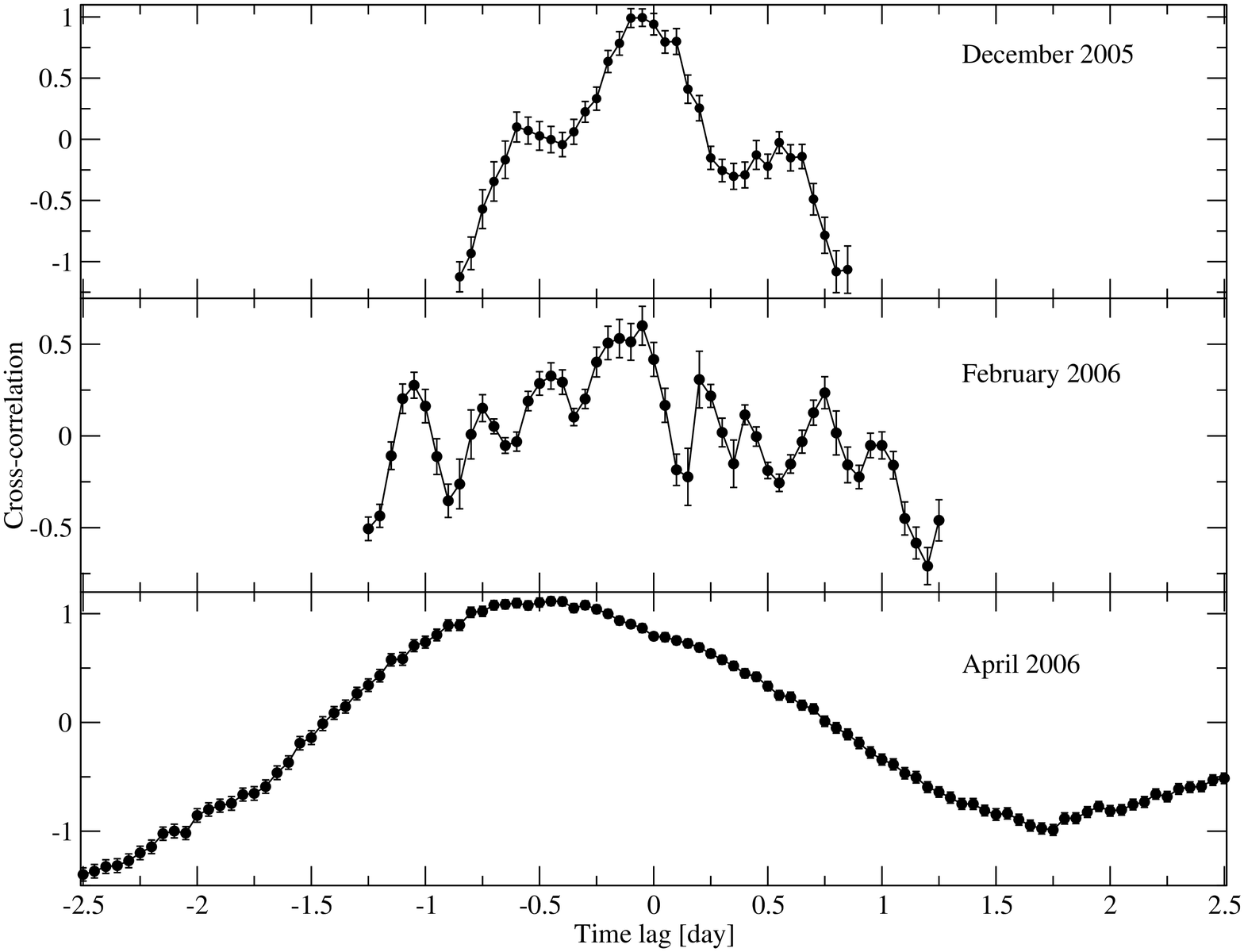}
  \caption{Cross-correlation functions of the light-curves of J\,1128+5925 
  at 2.70\,GHz and 4.85\,GHz for three different epochs. A negative time-lag 
  indicates that the lower frequency data are leading. The cross-correlation 
  was calculated using the method of \cite{corr_method}.}
\label{fig:corr}
 \end{center} 
\end{figure}

In Figure \ref{fig:freq}, simultaneously measured multi-frequency light-curves are displayed.
A correlation between the  2.70\,GHz and 4.85\,GHz data is obvious. This is 
consistent with the picture that the variations at both frequencies are due to scintillation
in the same, weak regime. This correlation is quantified by the cross-correlation 
functions shown in Fig. \ref{fig:corr}. The cross-correlation functions reveal 
a (insignificant) time lag between the 2.70\,GHz and the 4.85\,GHz light-curve 
of ($0.05\pm0.05$)\,day in December 2005 and in February 2006.
In April 2006, the time lag is larger ($-0.50 \pm 0.05$)\,day, showing that 
the 2.70\,GHz light-curve is leading. 

The cross-correlation function of the 4.85\,GHz and 10.45\,GHz light-curves of May 2005 
does not show a significant peak (its maximum value is $0.4\pm0.2$).
This may be due to the sparseness of the data at 10.45\,GHz. For 
the observations in December 2005 the correlation is higher, peaking at a time lag
$0.00 \pm 0.05$\,day with a correlation coefficient of $\sim 0.8$.


The differences between the observed time lags may be caused by a frequency dependent
structure of the scattering medium or of the scintillating source.
It is possible that the axial ratio or the position angle (or both) of
the anisotropy shows a certain frequency dependence and at different
times of the year, the scintillation patterns are displaced relative to
each other. According to the anisotropic annual modulation 
scenario, the relative velocity vector points almost 
at the position angle of the major axis of the scintillation pattern in 
April at 4.85\,GHz. Therefore the variability was long at this time.
In December, however, the velocity vector was almost perpendicular to the 
velocity vector in April, pointing closer to the direction of the 
minor axis of the scintillation pattern. The different time lags 
may arise because the ellipticity of the scintillation 
pattern at 2.70\,GHz is larger than at 4.85\,GHz; the major axes at the two 
frequencies differ by more than the minor axes. 

The larger and better defined time lag in April 2006 might also reflect 
frequency dependent size variations in an evolving radio source. This would
require that the VLBI structure of the source changes on a time scale of
$\sim 0.5$\,yrs, which is not unreasonable.

\section{Discussion and Summary}  \label{concl}

J\,1128+5925 is a new highly variable IDV source, which varies on time-scales
of 0.2\,days to 1.7\,days with peak amplitudes of up to $\sim 30$\,\% and
a modulation index in order of 10\,\%.

Adopting the usual light-travel time argument and by this a source intrinsic 
interpretations of the variability, the measured time-scales lead to
brightness temperatures of $\sim 10^{19-20}$\,K. This requires Doppler factors 
of at least a couple of hundred to reduce the brightness temperature 
to the inverse Compton or to the equipartition limit.

The variability time-scale of J\,1128+5925 is different at the 
different observing epochs and varies systematically over the year. 
We determine the variability time-scale by three different methods, which
ensures the robustness of the effect. We explored
the possibility if these changes can be described by the so-called annual 
modulation scenario, where variability is caused by interstellar scintillation,
and where the scintillation time-scale is modulated by the orbital motion of the
Earth around the Sun. We found that such annual modulation fits the data, if
some anisotropy is included. Similar models were used to 
describe the annual modulation observed in the much more rapid scintillators 
J\,1819+3845 \citep{annual1819} and PKS\,1257-326 \citep{bignall_newest}.

Whether the anisotropy originates from the source or from the scattering 
plasma is at present not clear. VLBI observation (CJF survey) of J\,1128+5925
show a slightly elongated core-jet structure, which is oriented 
at a position angle of $\sim -(91...100)^\circ$. This is in notable agreement 
with the position angle of the major axis of the anisotropy, as derived from 
the fitting of the annual modulation ($\sim -90^\circ$). 
Since the VLBI observations were performed more than ten years before our 
IDV measurements, it is unclear if the orientation of the source is still the same.
Further VLBI monitoring is necessary to search for a possible correlation
between source structure, its orientation and the anisotropy parameters of
scintillation models.

The frequency dependence of the variations (modulation index decreasing 
with increasing frequency) is in agreement with the theory of weak ISS. 
Using the frequency dependence of the modulation indices, we derived a source 
size, which is  inversely proportional to the frequency. This is in agreement with 
the standard jet model usually applied for a flat-spectrum radio sources (AGN).

Our observations yield an upper limit for the transition frequency between weak 
and strong scattering of $2.70$\,GHz. Using the
scintillation length scale from the annual modulation models and the 
frequency dependent source size (the size is proportional to $\nu^{-1}$), we derived a lower limit 
to the distance of the scattering screen of $\geq 50$\,pc for an isotropic 
scintillation model and $\geq 100$\,pc for an anisotropic model.

VLBI observations of extragalactic radio sources yield source sizes
which often are sufficiently small for showing interstellar scintillation.
Such sources also should show an annual modulation in their variability pattern,
if it is caused by scintillation. However up to now, a seasonal cycle
has been unambiguously identified only in two sources, both of which
belong to the class of extremely fast IDV sources \citep[rapid scintillators][]{annual1819,bignall_newest}. 
In the case of the slower, classical type II IDV sources, the detection
of annual modulation is more difficult, since the variability is slower
and can be less well extracted from time limited data trains. Furthermore
source-intrinsic structural variations also exist in many AGN, on 
time-scales of months to weeks. This can smear out or even destroy a seasonal cycle.
This might be the case in 0716+714 \citep{bach_vsop}.
Variations of the intrinsic source structure or of the
scattering medium could also explain temporal variations in the IDV pattern, 
and perhaps even a sudden stop of such variability. Such an effect may 
be seen in B\,0917+624, for which an annual modulation was previously proposed,
\citep{0917annual1, 0917annual2}, but later could not be confirmed due to the
ceasing of IDV \citep{cease_0917_2, cease_0917}. In fact a VLBI monitoring 
performed in parallel, shows structural variations in the core region of
B\,0917+624, which may be related to the observed decrease of the 
IDV amplitude and prolongation of the variability time scale \citep{cease_0917_3, cease_0917_4}. 
We note that episodic IDV was also observed in PKS B\,0405-385 
\citep{0405annual} and similarly was interpreted as being either due to variations 
in the physical properties of the scattering screen or a morphological evolution of the source.

To investigate whether source intrinsic variations play a role in the variability of J\,1128+5925 
further VLBI observations have to be carried out. During the 1.5 years of observations,
the source experienced a $20$\,\% increase followed by a similar decrease 
in its flux density at 4.85\,GHz. This could indicate variations of the intrinsic
source structure. Future kinematic VLBI studies of J\,1128+5925 can also help 
to derive a more accurate estimate of the Doppler factor, and by this of 
the source size and the resulting ``quenched'' scintillation. Together with
the scintillation length-scale derived from the annual modulation model, 
this would give more stringent limits to the distance of to the scattering screen.

In principle one needs to observe the source regularly for a longer period of 
at least two or three years to unambiguously identify an annual cycle 
in the scintillation. The observations of J\,1128+5925 presented here cover 
just 1.5 years. Only two observations were carried out at the same time of year, 
with one year separation (2004 December and 2005 December). The variability
time-scales extracted from these observations are
in very good agreement with the predictions of an annual modulation model. 
A prolongation of the variability time-scale can also be caused by an increase 
of the size of the scintillating component in the source, 
for example because of the ejection of a new jet component. 
Therefore, observation of the source during the time when we expect slower 
variability (here: in April-May and in August-September) is highly desirable. 
Over-subscription and logistical constraints at large radio telescopes
make such observations difficult. It is therefore useful (and necessary) to
share the observational load between different telescopes. The combination
of multi-frequency observations at two complementary telescopes 
(Effelsberg \& Urumqi) provides an efficient way to observe and detect 
systematic variations in the intra-day variability of radio sources, 
which vary slower than the so called rapid scintillators.

\begin{acknowledgements}
K. \'E. G. wishes to thank S.-J. Qian, H. Bignall, T. Beckert and A. Roy for the helpful discussions and 
A.\,Kraus for his help in the observations with the Effelsberg radio telescope 
and for his comments and valuable discussion concerning the data reduction.
K. \'E. G. also wishes to thank to S. Bernhart, E. Angelakis, V. Impellizzeri, 
and A. Pagels for their help in the observations at Effelsberg.  
This work is based on observations with the 100\,meter telescope of the 
MPIfR at Effelsberg (Germany) 
and with the 25\,meter Urumqi telescope of the Urumqi Observatory, National Astronomical Observatories of the Chinese Academy of Sciences.
K. \'E. G. and N. M. have been partly supported for this research through a stipend from the International Max Planck Research School (IMPRS)
for Radio and Infrared Astronomy at the Universities of Bonn and
Cologne. X.\,H. Sun and J.\,L. Han are supported
by the National Natural Science Foundation (NNSF) of China (10521001)
\end{acknowledgements}

\bibliographystyle{aa}
\bibliography{ref}
	
\end{document}